\documentclass[%
 reprint,
 amsmath,amssymb,
 aps,
prl,
rmp,
superscriptaddress,
]{revtex4-1}
\usepackage{lipsum}
\usepackage{graphicx}
\usepackage{color}
\usepackage{dcolumn}
\usepackage{bm}
\usepackage{textcomp}
\usepackage{gensymb}
\usepackage{appendix}

\begin{document}

\preprint{APS/123-QED}


 

\title{Observation and modelling of Stimulated Raman Scattering driven by an optically smoothed laser beam in experimental conditions relevant for Shock Ignition}


\author{G. Cristoforetti}
\altaffiliation{**}
\affiliation{Intense Laser Irradiation Laboratory, INO-CNR, Pisa, Italy}%
\author{S. H\"uller}
\affiliation{Centre de Physique Th\'eorique CPHT, CNRS, IP Paris, Ecole Polytechnique, Palaiseau, France}%
\author{P. Koester}
\affiliation{Intense Laser Irradiation Laboratory, INO-CNR, Pisa, Italy}%
\author{L. Antonelli}
\affiliation{York Plasma Institute, Department of Physics, University of York, York, UK}%
\author{S. Atzeni}
\affiliation{Dipartimento SBAI, Università di Roma ‘La Sapienza’, Roma, Italy}%
\author{F. Baffigi}
\affiliation{Intense Laser Irradiation Laboratory, INO-CNR, Pisa, Italy}%
\author{D. Batani}
\affiliation{Université Bordeaux, CNRS, CEA, CELIA, Talence, France}%
\author{C. Baird}
\affiliation{STFC Rutherford Appleton Lab, Central Laser Facility, Didcot, England}%
\author{N. Booth}
\affiliation{STFC Rutherford Appleton Lab, Central Laser Facility, Didcot, England}%
\author{M. Galimberti}
\affiliation{STFC Rutherford Appleton Lab, Central Laser Facility, Didcot, England}%
\author{K. Glize}
\affiliation{Key Laboratory for Laser Plasmas (MOE), Shanghai Jiao Tong University, Shanghai, China}%
\author{A. H\'eron}
\affiliation{Centre de Physique Th\'eorique CPHT, CNRS, IP Paris, Ecole Polytechnique, Palaiseau, France}%
\author{M. Khan}
\affiliation{York Plasma Institute, Department of Physics, University of York, York, UK}%
\author{P. Loiseau}
\affiliation{CEA, DAM, DIF, F-91297 Arpajon, France}%
\affiliation{Universit\'e Paris-Saclay, CEA, LMCE, 91680 Bruyères-le-Ch\^atel, France}
\author{D. Mancelli}
\affiliation{Université Bordeaux, CNRS, CEA, CELIA, Talence, France}%
\author{M. Notley}
\affiliation{STFC Rutherford Appleton Lab, Central Laser Facility, Didcot, England}%
\author{P. Oliveira}
\affiliation{STFC Rutherford Appleton Lab, Central Laser Facility, Didcot, England}%
\author{O. Renner}
\affiliation{Institute of Physics, ELI Beamlines, Institute of Plasma Physics, Czech Academy of Sciences, Prague, Czech Republic}
\author{M. Smid}
\affiliation{Helmholtz-Zentrum Dresden-Rossendorf, 01328 Dresden, Germany}
\author{A. Schiavi}
\affiliation{Dipartimento SBAI, Università di Roma ‘La Sapienza’, Roma, Italy}%
\author{G. Tran}
\affiliation{CEA, DAM, DIF, F-91297 Arpajon, France}%
\author{N.C. Woolsey}
\affiliation{York Plasma Institute, Department of Physics, University of York, York, UK}
\author{L.A. Gizzi}
\affiliation{Intense Laser Irradiation Laboratory, INO-CNR, Pisa, Italy}%

\date{\today}

\begin{abstract}
We report results and modelling of an experiment performed at the TAW Vulcan laser facility, aimed at investigating laser-plasma interaction in conditions which are of interest for the Shock Ignition scheme to Inertial Confinement Fusion, i.e. laser intensity higher than $10^{16}$ ${\rm~W/cm}^2$ impinging on a hot ($T > 1$ keV), inhomogeneous and long scalelength preformed plasma. Measurements show a significant SRS backscattering ($\sim 4-20 \%$ of laser energy) driven at low plasma densities and no signatures of TPD/SRS driven at the quarter critical density region. Results are satisfactorily reproduced by an analytical model accounting for the convective SRS growth in independent laser speckles, in conditions where the reflectivity is dominated by the contribution from the most intense speckles, where SRS gets saturated. Analytical and kinetic simulations well reproduce the onset of SRS at low plasma densities in a regime strongly affected by non linear Landau damping and by filamentation of the most intense laser speckles. The absence of TPD/SRS at higher densities is explained by pump depletion and plasma smoothing driven by filamentation. The prevalence of laser coupling in the low density profile justifies the low temperature measured for hot electrons ($7-12$ keV), well reproduced by numerical simulations.  
\end{abstract}

\pacs{Valid PACS appear here}
\maketitle


\section{INTRODUCTION}

After the difficulties encountered in the National Ignition Campaign (NIC)\cite{LindlNIC}, conducted at the National Ignition Facility (NIF), which were partially overcome in following years along with the demonstration of net energy gain \cite{Hurricane2014, Hurricane2019},
the scientific community working on Inertial Confinement Fusion (ICF) is looking with renewed and increasing interest to Direct Drive (DD) laser fusion schemes \cite{Bodner}, which have undoubted advantages with respect to the Indirect Drive (ID) approach \cite{craxton2015}.
First, the efficiency of laser energy coupling with the plasma corona is significantly larger, requiring a lower laser energy for achieving fuel ignition.
Furthermore, the ID approach is intrinsically non symmetric, with laser beams overlapping at the entrance of the hohlraum and propagating over long plasmas before irradiating the internal hohlraum surface; this produces undesired plasma instabilities (e.g. Crossed Beam Energy Transfer(CBET)) and suprathermal or hot electrons (HE), on one side, and a non uniform X-ray irradiation of the capsule, on the other. A symmetric irradiation scheme appears therefore a necessary precondition for reducing long-scale implosion asymmetries and for achieving a higher control of laser plasma interaction.

Among the DD schemes, Shock Ignition (SI), proposed by Betti \cite{betti2007}, is presently one of the most promising, and is therefore investigated in many recent works. Here, the fuel is ignited by a strong shock driven by an intense laser spike ($\sim 10^{16}$ W/cm$^2$) at the end of the compression phase. The main advantages of the SI scheme are the lower implosion velocity during the compression stage, strongly reducing the risk of Rayleigh Taylor instabilities, and the higher gain, enabling the ignition at moderate laser energies, already available in facilities like the NIF and Laser Megajoule (LMJ) \cite{atzeni2014,batani2014physics}.
On the other hand, the interaction of the laser spike with the long scalelength plasma corona surrounding the precompressed pellet - at intensities which are an order of magnitude higher than those envisaged in the classical DD scheme - results in an outburst of parametric instabilities, driven in a strongly non-linear regime that is not yet fully understood.
While the success of the original DD scheme requires a good comprehension and a full control of Stimulated Brillouin Scattering (SBS) and Two Plasmon Decay (TPD) instabilities, Shock Ignition scheme makes the scenario more tricky, involving also the onset of Stimulated Raman Scattering (SRS) and a boost of laser filamentation. Furthermore, fully kinetic Particle In Cell (PIC) simulations show the relevant competition between different instabilities and the strongly non-linear character of their growth. The former issue includes competition between instabilities driven in different plasma regions, i.e. by pump-depletion mechanisms, but also between instabilities driven in the same region, due to their different growth rates or damping. The non linear character of Laser-Plasma Interactions (LPI), on the other hand, involves the modification of the dispersion relation for plasma waves which are sufficiently intense, for example due to the electron trapping in the electron plasma waves (EPW), resulting in a shift of the plasma frequencies and in the consequent change of the instability growth rate. 

Reaching a detailed comprehension of LPI in the SI regime is important, not only because a large fraction of laser energy can be diverted out of the plasma via scattered light (SBS and SRS), therefore increasing the laser energy requirements, but also because some instabilities (SRS and TPD) result in the generation of HE, which propagate through the compressed pellet and can affect the shock strength and preheat the fuel, thus preventing fuel ignition. \\

Several recent experiments aimed at the investigation of LPI and HE generation in conditions relevant for Shock Ignition - i.e. laser intensities in the range $10^{15}-10^{16}$ W/cm$^{2}$, with wavelenght in the UV range, impinging on plasmas of a few keV temperature and of a few hundreds of microns density scalelength - have been reported. None of them, however, could meet all of these conditions simultaneously, because of laser energy limitations in the available laser facilities.
Results obtained in OMEGA, LULI and PALS facilities \cite{Baton_PRL, Theobald_pop, Cristoforetti_HPLSE, Cristoforetti_pop, Depierreux} suggested that SBS could be responsible for a large amount of scattered energy, with values reaching a few tens of percent of the laser incident energy.
Experiments also reveal the onset of SRS and TPD, but their relevance is strongly dependent on experimental conditions and their relative contribution is therefore more uncertain. 
The onset of TPD, driven at densities close to the quarter critical density, is usually observed  via detection of $\omega_0 /2$ and $3/2$ $\omega_0$ harmonics emission in the light scattered spectra, produced by the non linear coupling of incident laser light with EPWs driven by TPD. Its quantification is experimentally tricky as well as the determination of the amount and of the energy of the HE accelerated by the related plasma waves. However, while TPD is dominant in the traditional DD scheme, 2D PIC simulations suggest that its relevance could fall in SI conditions, because of SRS competition. Here, absolute SRS could prevail on TPD at densities close to the quarter critical density because of the higher growth rate, due to the dependence on the plasma temperature, while convective SRS at lower densities could also damp the TPD growth by pump depletion mechanisms.
Recent experiments at PALS carried out with $1\omega$ irradiation at $\approx 10^{16}$ W/cm$^2$ suggest that TPD is driven at early times, during the interaction of the leading part of the laser pulse, while it is successively damped, probably due to pump depletion caused by the onset of convective SRS at lower plasma densities \cite{Cristoforetti_HPLSE}.
In typical exploding-foil experiments, in fact, SRS is driven at later times of interaction, when the plasma scalelength has become sufficiently large, and is convectively amplified at densities well below the quarter critical density, close to the Landau damping cutoff determined by the plasma temperature ($k_{epw}\lambda_D\approx 0.3$). Very few experiments \cite{Baton, Depierreux, Montgomery2}, however, explored LPI at laser intensities close to $10^{16}$ W/cm$^{2}$ together with plasma density scalelength higher than $200$ $\mu$m, as envisaged in the SI scheme, where the non-linear character of SRS is expected to be strong. These works show that SRS is driven at very low densities, well below the Landau cutoff limit $k_{epw}\lambda_D>0.3$, where Landau damping is expected to severely reduce the instability growth rate. A full understanding of these observations is also made complex by the relevance of kinetic effects, due to the electron trapping into the EPW, which affects its dispersion relation\cite{MoralesONeil,Dewar,DewarLindl,RoseRussellPoP2001}, resulting in a reduced damping of the EPWs and/or in a shift of the resonance conditions \cite{Vu, Strozzi2007, Berger2013, Seaton, Spencer}.
A correct comprehension of this process is particularly relevant for SI, since EPWs driven at low densities are expected to generate very low energy HE ($\leq15$ keV), which could be beneficial for amplifying the shock pressure and would be unable to preheat the fuel to performance degrading levels.

In the present paper, we describe the results obtained in an experiment aimed at investigating the LPI of a laser pulse focused at an intensity of $\sim10^{16}$ W/cm$^{2}$ on a long pre-formed plasma, reaching a gradient scalelength of $\sim 400 \ \mu$m. Experimental results, showing the onset of SRS in strongly kinetic regime ($k_{epw}\lambda_D= 0.3-0.5$), are presented and discussed also in view of analytical and Particle In Cell simulations.

\section{EXPERIMENTAL SET-UP}

 The experiment was carried out at the Vulcan Laser in the Target Area West (TAW), at the Rutherford Appleton Laboratory. Four heating beams ($E=250$ J, $\lambda=1.053$ $\mu$m, $\tau=2.9$ ns) were focussed on a multilayer foil target by an $f\#/11$ optics to a $FWHM=570\times800$ $\mu$m$^2$ spot on the target surface to form an extended long scalelength plasma. The large spot size was conceived to produce a 1D plasma expansion in the interaction region and keep the intensity on the target, here $I=3\cdot10^{13}$ W/cm${^2}$, well below the threshold for the onset of parametric instabilities. The beams, smoothed by Random Phase Plates (RPP), were set at $\pm7^{\circ}$ and $\pm25^{\circ}$ to the horizontal and vertical axes, respectively.

\begin{figure}
\includegraphics[width=0.48\textwidth]{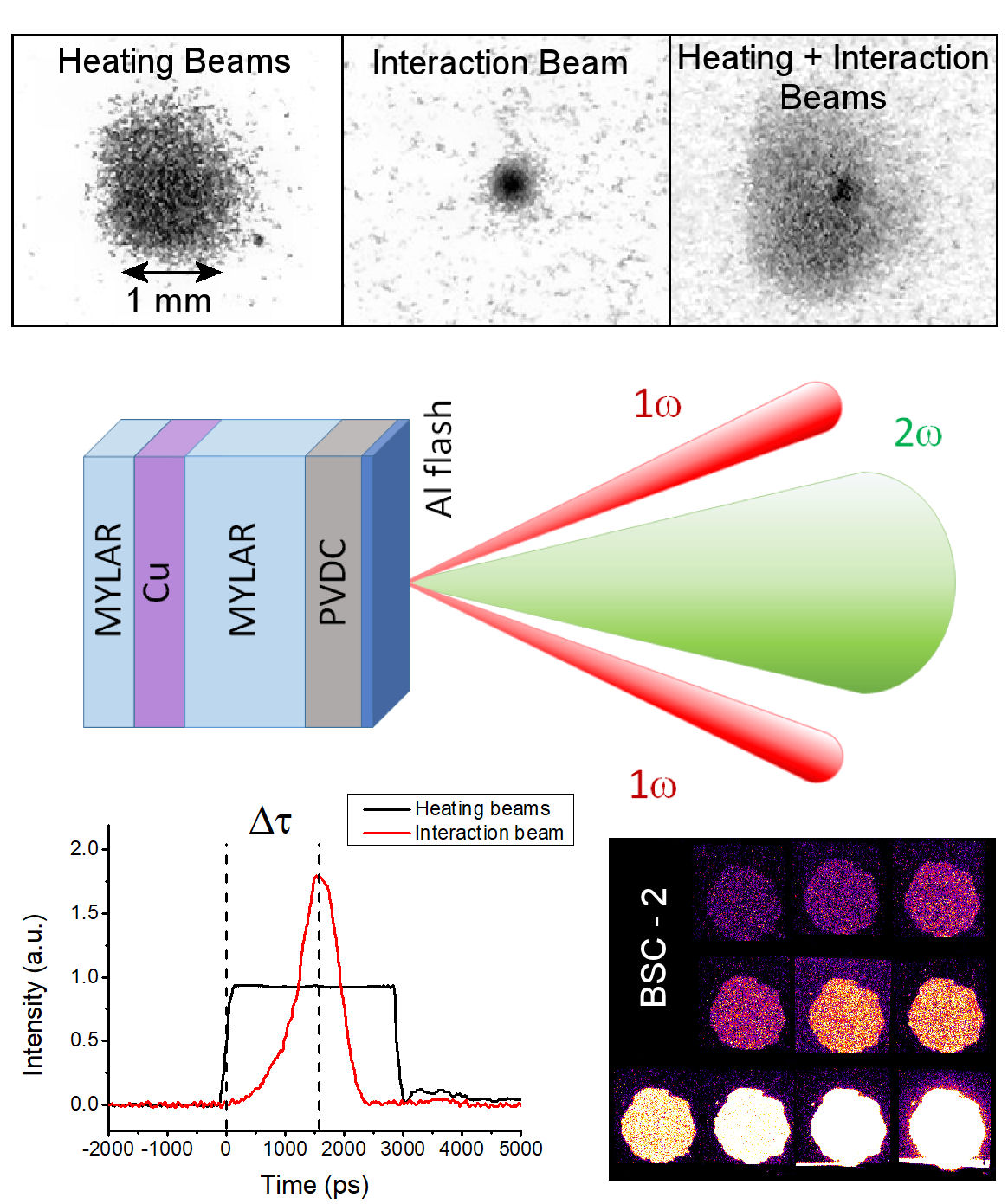}
\caption{\label{setup} Top, pinhole camera images of hard X-ray emission from plasmas induced by the heating pulses, by the interaction pulse and by all the beams. Middle, sketch of the beam configuration in the vertical plane and target structure. Bottom-left, timing of the laser beams (relative intensity is arbitrary). Bottom-right, image plates signal, obtained in calibration shot on a copper targets, acquired by the Bremsstrahlung Cannon looking at the rear side of the target; the intensity is proportional to PhotoStimulated Luminescence (PSL).}
\end{figure}
 
 The interaction beam ($E_{max}=100$ J, $\lambda=527$ nm, $\tau=700$ ps), smoothed by a RPP, was focussed normally (relative to the original target plane) on the preformed plasma by a $f\#/2.5$ lens to a $FWHM\approx24\times31$ $\mu$m$^2$ spot. The superposition of heating and interaction beams could be checked \emph{a posteriori} by a pinhole camera imaging of X-ray emission, filtered by a $6.5$ $\mu$m Al foil, as shown in Fig.\ref{setup}. The intensity on the target, calculated taking into account the pulse energy measured shot by shot by means of a calibrated calorimetric line, was in the range $(1-2)\cdot10^{16}$ W/cm${^2}$.
 Because of the long scalelength preformed plasma (see hydrosimulations below) we varied the focal position $\Delta x_{foc}$ of the interaction beam with respect to the original target surface position in the range from $-250$ $\mu$m to $+50$ $\mu$m, where the negative sign indicates that the laser waist is located before the target.
 The time delay $\Delta\tau$ between the rise front of the heating pulses and the peak of the interaction pulse (see Fig.\ref{setup}) was varied from 0.6 ns to 3.2 ns, with the aim of exploring different density gradients of the plasma at the time of the main pulse interaction.

 Multilayer targets were used during the experiment. The laser beams impinged on a 12 $\mu$m layer of PVDC $(C_2H_2Cl_2)_n$, mimicking the low-density ablation layer of an ICF capsule, over which a 100 nm thin film of aluminum was deposited to prevent the laser light to penetrate into the target in the early stages of interaction. The chlorine ions, present in PVDC, allowed the plasma temperature  to  be  measured  via  high-resolution X-ray spectroscopy. A Mylar layer of thickness varying between 0 (no layer) and 100 $\mu$m was located after PVDC and was followed by a 10 $\mu$m Cu tracer layer for detecting HE via $K_{\alpha}$ spectroscopy. Different values of Mylar thickness were used with the aim of controlling the amount of HE reaching the Cu layer to investigate their energy.
 A final 15 $\mu$m Mylar was located after the Cu tracer layer, with the scope of reducing the effect of HE refluxing on the $K_{\alpha}$ intensity. 

 Two high resolution X-ray spectroscopic diagnostics combining spectral and one-dimensional spatial resolution have been implemented. The Cu K$\alpha$ line emission was studied using an X-ray spectrometer equipped with a quartz (233) crystal, spherically bent to a radius of 150 mm, and protected by kapton (13 µm) and mylar (10 µm) foils. The spectrometer was set to look at the target at an angle of 11.8$^{\circ}$ from the normal and covered a spectral range from 1.39 \AA \ to 1.62 \AA, with a spatial demagnification of 0.34. Spectra were recorded using imaging plates BAS-MS (IP) and digitized by a Fuji scanner at a pixel size of 50×50 $\mu$m$^2$.
The macroscopic parameters of the plasma corona were studied via analysis of H- and He-like Cl spectra emitted from the PVCD coating. The second spectrometer was equipped with a spherical mica crystal with the bending radius of 150 mm, protected by 13-$\mu$m-thick kapton foil. The instrument covered the spectral ranges from 4.16 \AA \ to 4.53 \AA \  and from 3.33 \AA \ to 3.63 \AA, diffracted in the 4$^{th}$ and 5$^{th}$ diffraction order, respectively and was set to look at the target at an angle of 17.5$^{\circ}\pm$ 0.5$^{\circ}$ vs the target surface. The spectra were again recorded on BAS-MS IP, digitized and corrected with respect to the wavelength dependent crystal reflectivity and filter transmission. The wavelength calibration was based on the ray-traced dispersion relation and tabulated wavelengths of the dominant X-ray lines.
 
 In addition to $K_{\alpha}$ detection, HE were also characterized by measuring the Bremsstrahlung X-ray emission by means of two spectrometer "cannons", looking at the front (BSC-1) and at rear (BSC-2) sides of the target at angles of 45$^{\circ}$ to their respective normal axes.
 Bremsstrahlung cannons were designed by relying on K-edge and differential filtering, with the atomic number Z of the filters increasing from Al to Pb, and using images plates (IPs) as detectors \cite{Chen}. The stack of filters and IPs was housed in a lead shielding box and combined with a collimating system and a magnet for deflecting high energy electrons. An example IP scan from the BSC-2 cannon is shown in Fig.\ref{setup}.
 
 Laser-Plasma Instabilities were investigated by means of calorimetry and time-resolved spectroscopy of light backscattered in the cone of the focussing optics of the interaction beam. Light was collected behind the last turning mirror of the laser transport line, separated in four different channels, and sent to two calorimeters and two time-resolved spectrometers.
 Spectral filters were placed in front of the calorimeters to select light scattered by SRS ($\omega < 0.8 \,\omega_0$) and wavelengths close to $\lambda = 527$ nm including SBS and laser backscattering ($0.8\,\omega_0< \omega < 1.5 \,\omega_0$). The measured energy, combined to an accurate measurement of the spectral transmissivity of the optical line, yielded the plasma reflectivity in these spectral ranges.
 Time-resolved spectrometers ($\Delta t_{min}\approx 7$ ps), consisting of monochromators (Acton SP2300i) coupled to fast Streak cameras (Hamamatsu C7700 and C5680), were devoted to measure the scattered light in the whole spectral range going from $\omega_0/2$ to $3/2$ $\omega_0$, including both SRS and half-harmonics derived from the coupling of laser light with EPWs driven by TPD. The two spectrometers were equipped with gratings of 300 l/mm and 600 l/mm and were coupled to 512x512 and 1280x1024 pixels CCD, resulting in spectral ranges of $\sim 280$ nm and $\sim 150$ nm, respectively. A laser pick-off was sent to the streak photocathode and used as a fiducial signal for the absolute time calibration of SRS emission.
\section{INTERACTION CONDITIONS}
The interaction conditions of the main laser pulse with the plasma corona were modeled by using the DUED\cite{DUED} hydrodynamic code. 2D maps of density and temperature were simulated for the cases when no heating beams were used and when the main beam was delayed by $\Delta\tau=[0.6 \div 3.2]$ ns with respect to the rising front of the heating beams.
\begin{figure}
\includegraphics[width=0.48\textwidth]{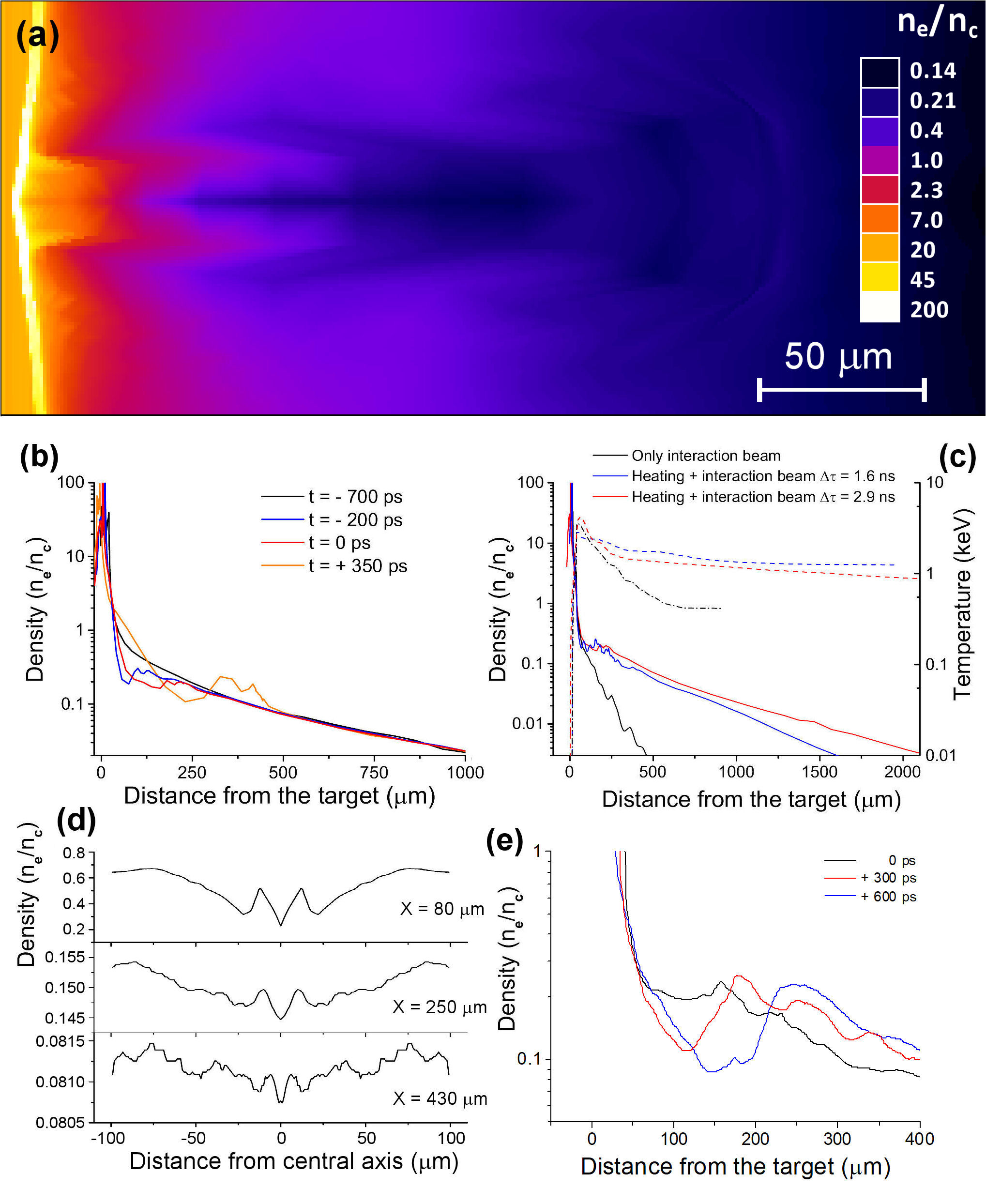}
\caption{\label{DUED} Density and temperature profiles obtained from hydrodynamic simulations carried out with the DUED code. (a) 2D map of electron density in the high density region taken at the peak of the interaction pulse, in case of delay time $\Delta\tau=2.9$ ns; (b) longitudinal profiles of electron density at different times of interaction in case of delay time $\Delta\tau=2.9$ ns; (c) longitudinal profiles of electron density and temperature taken at the peak of the interaction pulse for delay times $\Delta\tau=1.6$ ns (blue lines), $\Delta\tau=2.9$ ns (red lines) and for the case where heating beams are not used (black lines); (d) transverse density profiles taken at different distances from the target surface in the same conditions as (a); (e) temporal evolution of the dip in the density profile.}
\end{figure}

When only the interaction beam is used, the density profile at the laser peak shows a change of slope around $n\approx 0.2$ $n_c$ (Fig.\ref{DUED}c), resulting in a density scalelength $L_{\nabla}=n/ (dn/dx)\approx90$ $\mu$m at lower densities and $L_{\nabla}\approx25$ $\mu$m for higher densities. The region around $n=0.2$ $n_c$ absorbs most of the laser energy and therefore shows the maximum of plasma temperature, which rapidly falls in more rarefied regions of the plasma.

When heating beams are used, the main pulse impinges on a long preformed plasma corona with a density scalelength of several hundreds of $\mu$m, increasing with the time delay $\Delta\tau$ between the beams. Values of $L_{\nabla}\approx 380$ $\mu$m and $L_{\nabla}\approx300$ $\mu$m for a delay $\Delta\tau=1.6$ ns, and of $L_{\nabla}\approx450$ $\mu$m and $L_{\nabla}\approx275$ $\mu$m for a delay $\Delta\tau=2.9$ ns, are obtained at densites $n=0.04$ $n_c$ and $n=0.1$ $n_c$, respectively. The intense beam heats the plasma along its path, digging a hot low-density channel with a transverse size of the order of the laser waist, as shown in Figs.\ref{DUED}a and \ref{DUED}d. The channel is weakly visible in the rarefied regions, but becomes deeper at densities larger than $n=0.15$ $n_c$. The strong absorption of laser light at $n>0.15$ $n_c$ produces a hot rarefied plasma bubble, digging progressively a dip in the longitudinal profile of electron density and producing a steepening at densities $n>0.2$ $n_c$ (Fig.\ref{DUED}e). The bubble propagates toward lower densities at successive times, producing a modulation in the density profile, down to $n\approx 0.1$ $n_c$. This strongly affects the interaction conditions of the main pulse at densities higher than $n\approx 0.1$ $n_c$, which are self-consistently determined by the main beam itself. 

In contrast, the interaction conditions at densities lower than $n\approx 0.1$ $n_c$ are mainly determined by the heating beams, and are characterized by a plasma with a temperature of $\sim 1-1.2$ keV and an exponentially decreasing density profile.

\begin{figure}
\includegraphics[width=0.48\textwidth]{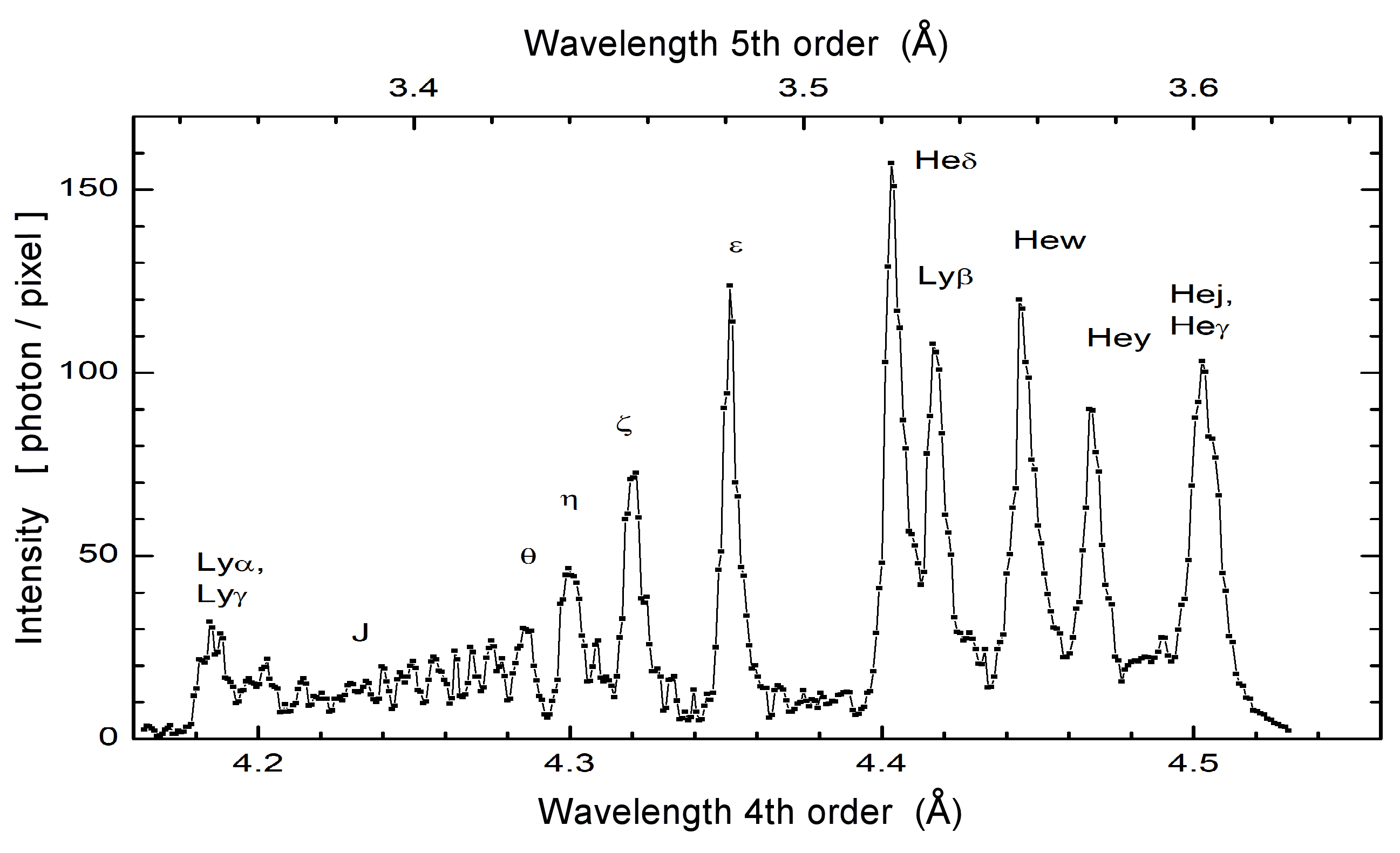}
\caption{\label{Chlorine}X-ray spectrum of K-shell Chlorine emission from H- and He-like atoms. The measured spectrum includes the contributions of the 4$^{th}$ and of the 5$^{th}$ diffraction order of the crystal.}
\end{figure}

Spatially resolved X-Ray chlorine spectroscopy allowed us to calculate the plasma temperature in different regions of the plume. A typical spectral lineout, referring to the K-shell emission from H- and He-like dopant Cl atoms at 800 $\mu$m from the target surface, is shown in Fig.\ref{Chlorine}. It includes well resolved lines in spectral ranges 3.3-3.7 {\AA} and 4.1-4.6 {\AA}, given by the 5$^{th}$ and by the 4$^{th}$ crystallographic orders, respectively. Temperature is here obtained by the ratio of Ly$_{\beta}$ and He$_{\delta}$ lines\cite{Smid}, after a comparison with synthetic spectra calculated with the PrismSpect code\cite{PRISMSPECT}.
In the region of interest for parametric instabilities, i.e. in the range $0.03-0.12$ $n_c$, the temperature retrieved by Cl X-ray spectra is $\approx$ 700-800 eV, a value significantly lower than the value obtained by hydrodynamic simulations.
This discrepancy could be due, on one hand, to the time-integration and to the spatial integration in the transverse direction of the spectral measurements, leading to an underestimation with respect to the local conditions on the laser axis. On the other hand, the use of the nominal laser intensity in the hydrodynamic simulations, thus neglecting the energy scattered by parametric instabilities, could also produce an overestimation of the plasma temperature.

Local conditions of interaction, i.e. plasma temperature and density gradient, could be here also affected by the onset of filamentation, driven by the self-focusing of the speckles produced by the RPP. 
The relevance of filamentation can be estimated by considering a density/temperature perturbation size of the order of the speckle size $l_{\perp}=1.2\lambda f_{\#}=1.6\,\mu m$, where $f_\# =2.5$ is the f-number of the focusing system. A hot spot in a laser speckle is stable to self-focusing if the spatial growth gain $G=\kappa_g \cdot l_{\parallel}$, where $\kappa_g$ and $l_{\parallel}$ are the spatial grwoth rate and the speckle length, respectively, is less than unity \cite{Berger}.
By considering a gaussian-shaped speckle, its length $l_{\parallel}$ can be estimated by integrating along the $x$ longitudinal direction, $l_{\parallel}\simeq\int_{-\infty}^{\infty} dx/[1+(x/L_R)^2] = \pi L_R$, where $L_R \approx 2.8 f_\#^2 \lambda_0$ is its Rayleigh length \cite{HullerLPB2010}, obtaining $l_{\parallel} \simeq 29$ $\mu m$.
According to the local conditions described above, the critical power for ponderomotive self focusing is 930 MW, 360 MW and 280 MW at $n= 0.04, 0.1$ and $0.25$ $n_c$, respectively. These values are larger than the average power in a speckle, which is $\approx250$ MW, suggesting that self focussing is driven only in most intense speckles.
At densities around $n=0.04$ $n_c$, which are relevant for the present experiment (see below), self focussing is driven in speckles with intensities $I>3.5\langle I\rangle$, where $\langle I \rangle$ is the intensity of the laser envelope. This threshold also accounts for the reduction due to non local electron heat transport\cite{Epperlein,brantov}; this follows from the fact that the electron mean free path $\lambda_e\approx20$ $\mu m$ is here much longer than the temperature perturbation size, $l_{\perp}\sim 1.6$ $\mu m$, resulting in a reduced capability to dissipate the temperature gradients.
Considering the experimental conditions, we estimate a number of speckles of $\sim16000$ in the focal volume; assuming an exponential intensity distribution as given by RPP smoothing model \cite{rose93}, this implies that intensities up to 8-10 $\langle I\rangle$ are reached in the most intense speckles.

The validity of the above estimation can be corroborated by calculating the spatial growth rate $\kappa_g$ of filamentation, including ponderomotive effects and thermal correction, and the net growth in a speckle length $l_{\parallel}$. According to \cite{Epperlein}, $\kappa_g=0.17$ $\mu m^{-1}$ in a speckle with $I=3.5\langle I \rangle$, which implies that $\kappa_g \cdot l_{\parallel}>1$, i.e. that the instability can significantly grow into the length of a speckle.
\section{EXPERIMENTAL RESULTS}
\subsection{Laser Plasma Instabilities}
Light backscattered at $\lambda \approx 527$ nm consisted of $15-35\;\%$ of laser energy, with no clear dependence, in the explored range, on laser intensity or time delay between heating and interaction beams; this value fell to $7-8\;\%$ when the heating beams were not used. It is worth to remark that the spectral resolution of the diagnostics did not allow to distinguish between SBS and laser light backscattered by the plasma.
\begin{figure}
\includegraphics[width=0.48\textwidth]{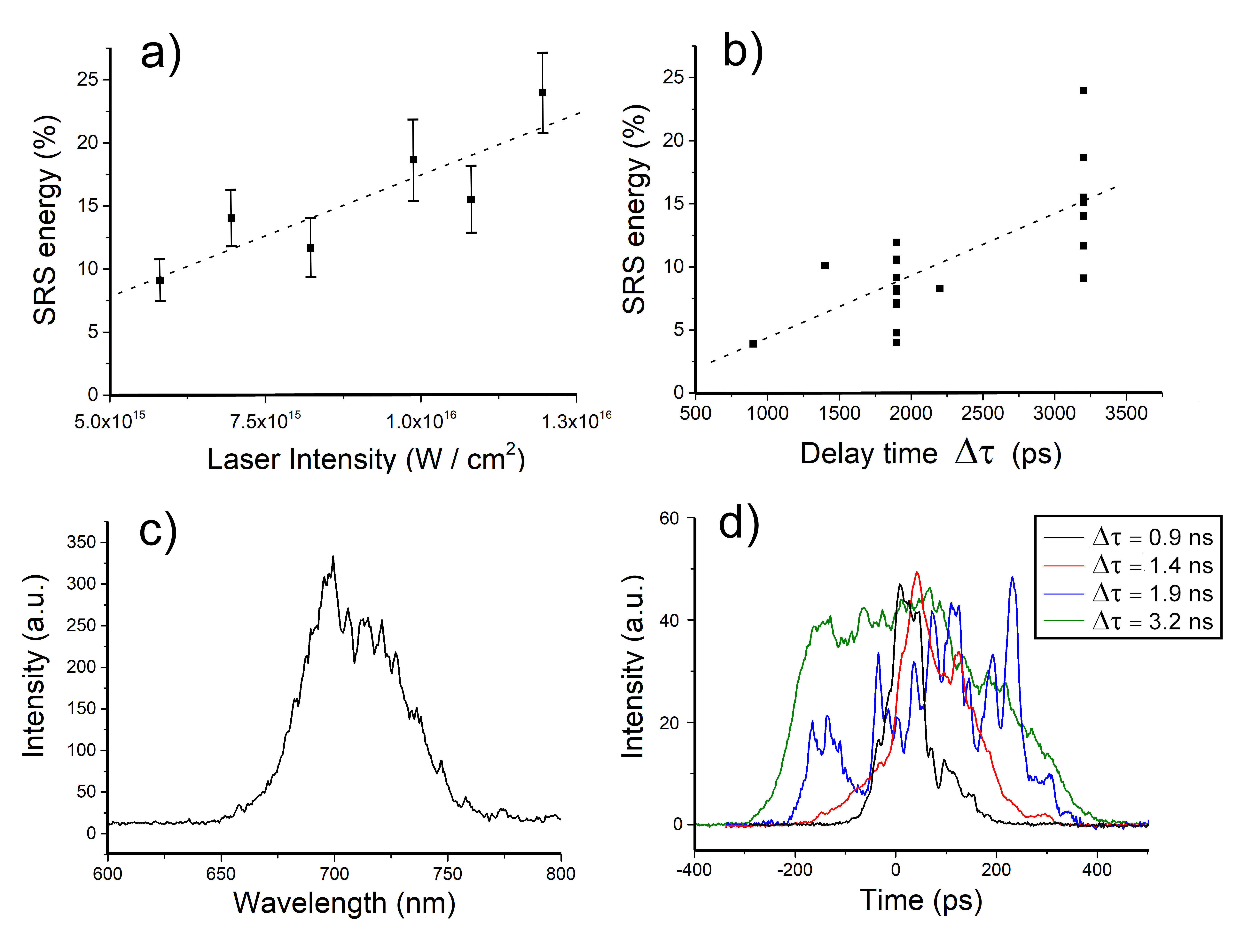}
\caption{\label{SRS} Experimental results. a) SRS reflectivity versus laser intensity for a set of selected data with fixed values of $\Delta\tau\;=\;3.2$ ns and $\Delta x_{foc}\;=\;-150\; \mu$m;  b) SRS reflectivity versus the delay time $\Delta\tau$ between heating and interaction beams; c) Typical SRS spectrum. d) Time profile of SRS light in shots with different delay time $\Delta\tau$. In subplot a) error bars of 20\% have also been reported for reference.}
\end{figure}

The SRS calorimeter measured no signal in the shots where only the heating beams or only the interaction beam were used. The former result suggests that no spurious signal produced by the heating beams affects the calorimetric measurements in shots with both heating and interaction beams. The latter observation can be explained by the fact that the plasma produced by the interaction beam is too steep to drive convective SRS. Considering the limits of detection of the optical line, this means that backscatter obtained by using only the interaction beam was lower than 0.5 $\%$ of laser energy.
In the shots when both heating and interaction beams were used, a clear SRS signal was detected, with the fraction of the interaction beam energy backscattered by SRS varying in the range $4-20\;\%$ of the interaction beam energy. The value was clearly dependent on the laser intensity, on the focal position $\Delta x_{foc}$ and on the time delay $\Delta\tau$ between heating and interaction laser pulses. The dependence on laser intensity is shown in Fig.\ref{SRS}a, where an homogeneous set of shots with fixed values of $\Delta\tau\;=\;3.2$ ns and $\Delta x_{foc}\;=\;-150\; \mu$m is selected.
The effect of $\Delta\tau$ can be observed in Fig.\ref{SRS}b, where however a large variability of the SRS energy is visible for each time delay, due to the included range of laser intensities and of focal positions; this last parameter affects in turn the local laser intensity at the density where SRS is driven.
The increasing trend shown in Fig.\ref{SRS}b can be ascribed to the progressive larger value of the density scalelength with the time delay, as shown in Fig.\ref{DUED}c, and is a clear hint that SRS growth has a convective character; the trend may also be affected by the progressive reduction of plasma temperature in the region of interest, and therefore of Landau damping of EPW, with the time delay, as also visible in Fig.\ref{DUED}c.
\begin{figure}
\includegraphics[width=0.48\textwidth]{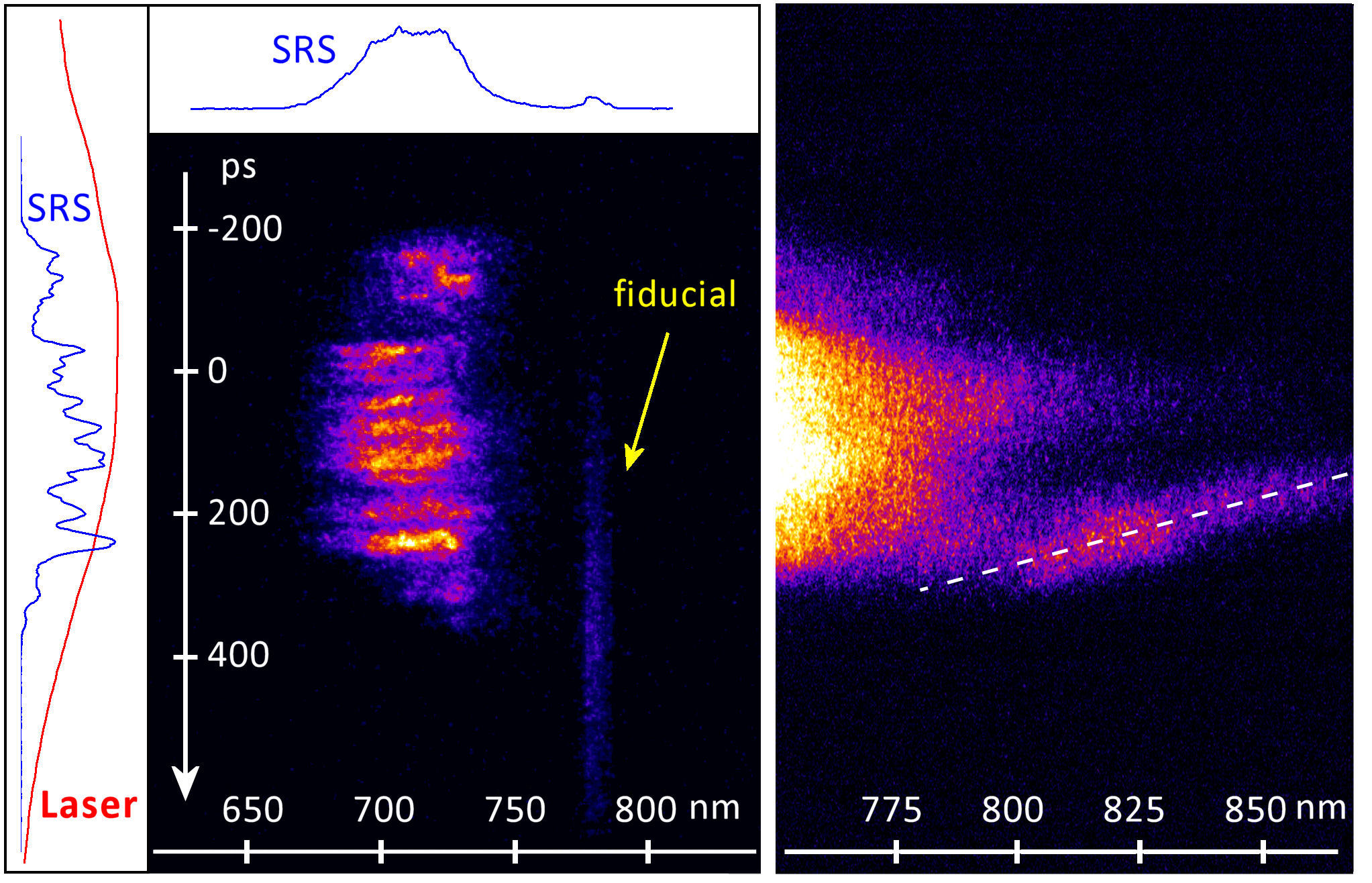}
\caption{\label{Streak} Time-resolved SRS spectra acquired in two different shots by the Hamamatsu C7700 (on the left) and C5680 (on the right) streak cameras. Time resolution is 18 ps. The laser fiducial in the left image allowed to determine the absolute timing of the interaction laser peak. Left and right panels show the SRS spectra scattered at low and high plasma densities, respectively. The dashed line shows the spectral shift of SRS scattered light with time at densities around $0.1$ n$_c$.}
\end{figure}

Detailed information about the timing of parametric instabilities and the plasma density where they are driven can be inferred by time-resolved spectroscopy of backscattered light.
In none of the laser shots, a clear $\omega_0/2$ or $3/2$ $\omega_0$ signal was detected, although different gratings, filtering and timing configurations were attempted. The absence of half-harmonics suggests that TPD and absolute SRS are here not driven, differently from other experiments carried out at similar laser intensities. Time-resolved SRS spectra were detected in all the shots where both heating and interaction beams were used. A typical spectrum is shown in Fig.\ref{Streak}. The strongest signal was detected in the spectral region ranging from 680 nm to 730 nm, as shown in Fig.\ref{SRS}c, corresponding to a plasma density spanning from 0.03 to 0.07 $n_c$. Measurements show that emission in this spectral region is not affected by the laser intensity nor by the time delay $\Delta\tau$, and
also clearly show that SRS is driven in successive bursts. As visible in Fig.\ref{Streak}, in each burst SRS light is emitted at the same time (within time resolution of the spectrometer of 7 ps) in a large spectral region, going from 680 nm to 730 nm, approximately, corresponding to the full 0.03-0.07 $n_c$ density range.

The central SRS wavelength does not shift with time, indicating that local conditions where SRS is driven are stationary. This suggests that local pre-formed plasma conditions, as determined by the heating beams, namely the temperature and density profiles in regions $n<0.1$ $n_c$, are not affected significantly by the interaction beam as expected from hydrodynamic simulations.
The SRS signal is approximately peaked 0-100 ps after the laser peak and its duration increases with the delay $\Delta\tau$, as shown in Fig.\ref{SRS}d.

In a few shots, an additional SRS signal is observed at wavelengths larger than 780 nm, as shown in the right image of Fig.\ref{Streak}. This signal is much weaker than that observed at lower wavelengths, and consists at maximum of a few percents of the main SRS reflectivity. Here SRS begins $\sim$200-300 ps after the laser peak and lasts for a few hundreds picoseconds. This signal is peaked at wavelengths moving with time from $\lambda \approx$870 nm to $\lambda \approx$780 nm; accounting for Bohm-Gross dispersion relation, this implies that SRS here progressively shifts from $n_e/n_c \approx 0.13$ to $n_e/n_c \approx 0.08$. As shown in Fig.\ref{DUED}b, the density profile in this range is strongly affected by the interaction of the main laser pulse and varies with time, which explains also the time-variation of SRS wavelength.
\subsection{Hot Electrons}
Energy and amount of HE were here retrieved by the Bremsstrahlung Cannons measurements (BSC).
All Image Plates were scanned after $\sim$20 minutes from exposure to reduce the uncertainties produced by the signal decay with time and corrected for the time fading as in Ref.\cite{Boutoux}. The signal was detected only in the first 5-6 IPs, despite the adoption of several filtering configurations and the addition of extra lead shielding for noise reduction.
\begin{figure}
\includegraphics[width=0.48\textwidth]{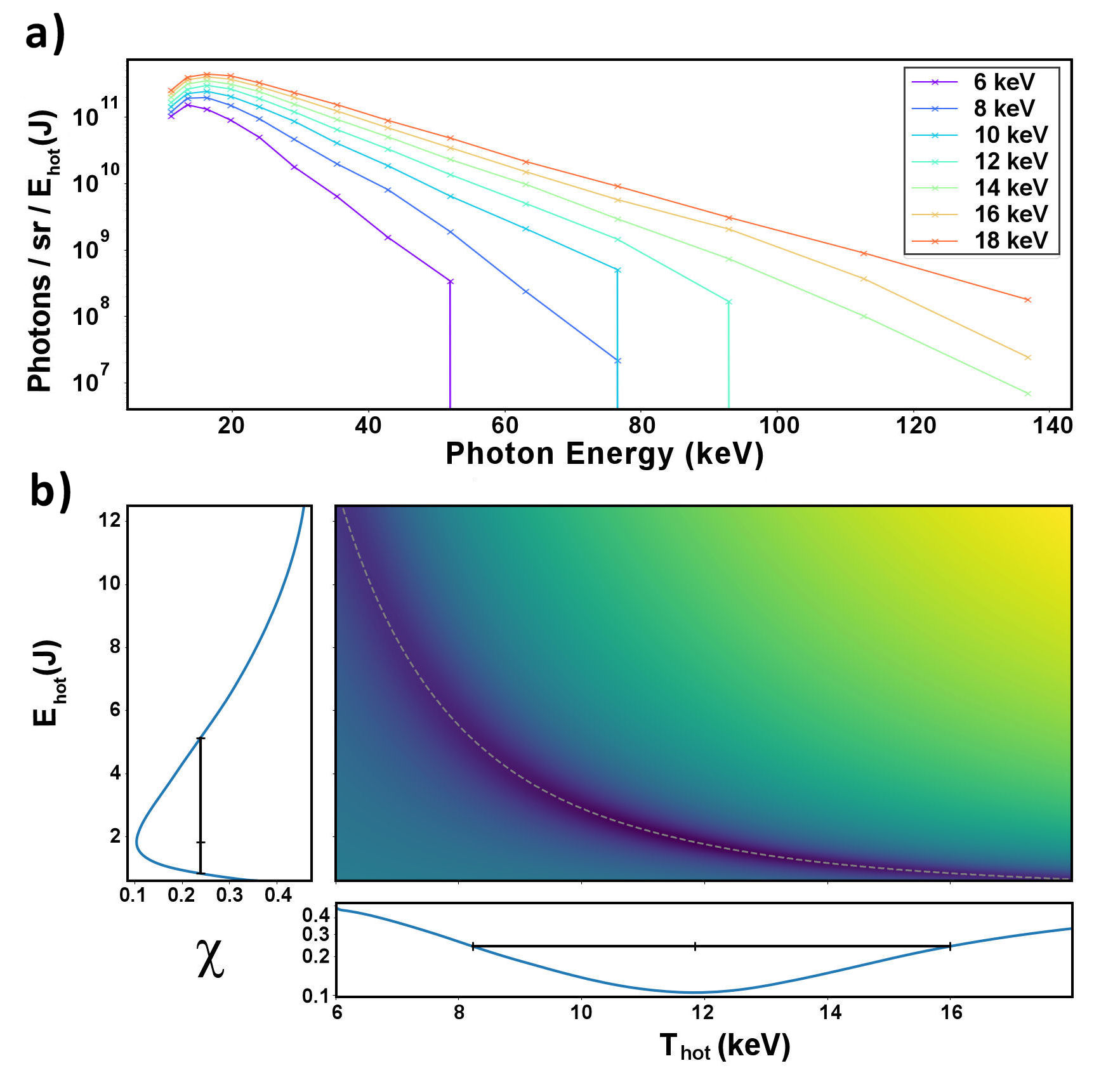}
\caption{\label{cannon} a) Synthetic Bremsstrahlung X-ray emission spectra obtained by HE populations of different $T_{hot}$ temperatures, resulting from Geant4 simulations. b) Typical heat map of $\chi^2$ for different ($T_{hot}$,$E_{hot}$) combinations; the gray line represents the locus of minimum chi squared, while the projections on the axes show that the optimal fit consists of $T_{hot}=12$ keV and $E_{hot}=2$ J.}
\end{figure}
The first IP was also discarded because it was expected to be affected by plasma self-emission.
On the other IPs, the signal was obtained by extracting the PSL from an ROI centred on the exposed part of the IP and by subtracting the background measurement taken from an unexposed IP. However, due to the significant non-uniformity of the background signal, a line-out was taken across each IP and top hat functions, including both a constant and linear background terms, were used to fit both signal and background regions.
The extracted data were fitted by synthetic signals produced with a combinations of a Geant4\cite{ALLISON} simulations for both the detector response and bremsstrahlung emission from hot electron propagating through a cold, unexpanded multilayer target, as shown in Fig.\ref{cannon}a. 
The injected electrons had a Maxwellian distribution of the form
\begin{equation}
  f(E,T_{hot})= \frac{2\sqrt{E}}{\sqrt{\pi} T_{hot}^{3/2}}\exp{(-E/T_{hot})}, \label{eq1} 
\end{equation}
with the hot electron temperature $T_{hot}$ ranging from 6 to 20 keV.

A residual sum minimisation analysis between experimental and synthetic signals was carried out by varying both the electron temperature $T_{hot}$ and their total energy $E_{hot}=3/2\; N_{HE}T_{hot}$, where $N_{HE}$ is the total HE number. A typical heat map of the value of $\chi^2$ for each ($T_{hot}$,$E_{hot}$) combination is reported in Fig.\ref{cannon}b, with the dark blue representing the lowest values and thus the best fitting. 
The projection of $\chi^2$ along the blue curve for each individual parameter allows to visualize the optimal ($T_{hot}$,$E_{hot}$) combination and to determine its uncertainty. 

The results obtained for both cannons revealed temperatures in the range 7-12 keV with an
uncertainty of $\sim20-30\%$. Total HE energy was spanning from 1 J to 7 J, corresponding to values of energy conversion efficiency of $\sim1-4\%$.

A low value of the HE temperature is in agreement with the results obtained by Cu K$_\alpha$ spectroscopy; also in this case the signals were very weak and visible only in a limited number of shots. The stopping range of 10 keV electrons in Mylar, calculated in Continuous Slowing Down Approximation (CSDA), is in fact $\sim2$ $\mu$m, much smaller than the thinnest layer of plastic used in the targets before the Cu tracer layer.
No clear correlation was found between the energy and the temperature of the HE with the laser intensity/energy nor with the SRS backscattered energy. Moreover, similar results were retrieved for the shots where heating and interaction beams were fired and for the shots where only the main beam was fired. 
These results suggest that multiple mechanisms are here responsible for the generation of low-energy HE, among them also the SRS driven at low densities is likely to contribute as its scattered light spectra are compatible with the measured HE energies, but no clear evidence of the role of SRS in the generation of HE was found. 
\section{Discussion}
Experimental results show that LPI occurs far from the critical density region, where SRS is predominantly driven at density $n\approx$ 0.04 $n_c$, a weak SRS is occasionally measured at densities close to $0.10$ $n_c$ and no instabilities taking place at the quarter critical density are observed. These features can be produced by the concurrence of several factors.
Firstly, both the high laser intensity and the long density scalelength of the plasma favour a strong SRS growth at low densities, as discussed below, producing a considerable amount of energy which is backreflected (and a corresponding amount of energy which is absorbed according to Manley-Rowe relations). This produces a significant pump depletion of the laser pulse before it reaches higher densities.
Moreover, at the laser beam intensities applied in our experiment we can expect that the more intense speckles, having a peak intensity up to 8-10 times higher than the average value, would be subject to self focusing and filamentation. Due to the short length of the speckles, resulting in speckle layers in the propagation direction of $\approx 30\ \mu$m, the dynamics of self focusing will inevitably lead to a spatial but also temporal incoherence further inside the plasma, practically after a two speckle layer. This process, provoking the so-called "dancing filaments" \cite{SchmittAfeyan}, leads to plasma-induced smoothing\cite{maximov2001,labaune2000,LoiseauPRL} that is known to prevent or strongly reduce the onset of SRS at higher densities in the plasma profile.
Finally, according to hydrodynamic simulations, a significant collisional absorption is expected in the region of densities $0.15-0.20$ $n_c$. 
The absence of $3/2$ $\omega_0$ emission could be therefore explained by the lower amount of laser light reaching the quarter critical density region, by the reduced coeherence of the beam, as well as by the local conditions in the $n_c /4$ region. According to hydrosimulations, in fact, the plasma temperature at $n_c /4$ is as high as 3-4 keV and the density profile is significantly steep, as shown in Fig.\ref{DUED}e, both these features resulting in a rise of the TPD threshold to $I_{TPD}\approx3\cdot10^{15}$ W/cm$^2$.
%
\subsection{SRS at low densities}
SRS is mainly driven in the electron density range, $n_e$, between 0.03 $n_c$ and 0.07 $n_c$, giving rise to Electron Plasma Wave (EPW) frequencies $\omega_{epw}\!=$ (0.22\ldots 0.29)$\omega_0$ and wave numbers $k_{epw}\!= $(1.75\ldots 1.63)$\omega_0/c$, respectively.
By considering electron temperatures in the range of 1\ldots 1.2 keV in the region of interest, the resulting Debye length values, $\lambda_D=(0.3\ldots 0.5)/k_{epw}$, indicate that EPWs should be under the influence of strong linear Landau damping.

In order to depict the physics of Stimulated Raman Scattering in the conditions of interest, in the next sections we try to disentangle various issues affecting the plasma response, tackling progressively (i) the effect of beam smoothing with the RPP, (ii) the relevance of kinetic effects and finally (iii) the role of the filamentation.
\subsubsection{Beam smoothing and role of laser speckles}
In order to account for the SRS driven in a multispeckles focal volume, as produced by the Random Phase Plate, we consider a simplified model consisting of SRS growth in independent laser speckles. The model, described in detail in the Appendix\ref{Appendix}, assumes that each laser speckle contributes incoherently to the backscattered light from SRS, according to its local intensity $I_{sp}$, where laser intensities of the speckles follow a probability distribution $f(u=I_{sp} / \langle I\rangle)$.   
SRS reflectivity is here calculated in each speckle according to the classical theory formulated by Rosenbluth\cite{Rosenbluth} for a convective growth in an inhomogeneous plasma, i.e. neglecting kinetic effects on the laser plasma coupling, and expressed by $R_{sp}(u)  = \varepsilon\ e^{g_0 u}$; here, $\varepsilon = I_{noise} / u \langle I\rangle$ stands for the ratio of the noise level to the speckle intensity, whose typical value for warm plasmas in laser plasma interaction is roughly $10^{-9}$ and $g_0$ is the amplification gain of a speckle at average laser intensity $\langle I\rangle$.
The model accounts for the saturation of SRS in the most intense speckles, levelling their response to a constant value of saturated reflectivity $R_{sat}$. 
The physics of the saturated regime can be very complex because of the concomitance of numerous non linear effects in intense speckles.
Saturation can be produced by the depletion of the incident flux into the speckle, or by non linear processes in the coupling process\cite{YinPoP2009,Spencer,RusselPoP99,CohenPoP2001,VTT-FuchsPoP2001,Ghizzo,Albrecht,Strozzi2007}, strongly limiting the amplitude of the plasma wave responsible for the laser light scattering. 
In our model, $R_{sat}$ denotes, for simplicity, a time-average value, smoothing a possible bursty SRS behaviour\cite{YinPoP2013}. According to previous experimental and simulation results, it can be estimated of the order of $0.4-0.5$.

According to the model, the overall SRS reflectivity can be written as
\begin{equation}
\langle R \rangle= \varepsilon \int_{0}^{u_{sat}}\!\! u\ e^{g_0 u} f(u)\ du + R_{sat} \int_{u_{sat}}^{u_{max}}\!\! u\ f(u)\ du  ,\label{eq_refl}
\end{equation}
where the two terms express the contribution from speckles where SRS grows in the linear or in the saturated regime, respectively. In Eq.(\ref{eq_refl}), $u_{max}= I_{max}/\langle I\rangle$ represents the highest intensity achieved in the speckle ensemble, while $u_{sat}= g_0 \log (R_{sat}/\varepsilon) \simeq (20 -\log R_{sat})/g_0$ represents the intensity for which saturation occurs. 
\begin{figure}
\includegraphics[width=0.48\textwidth]{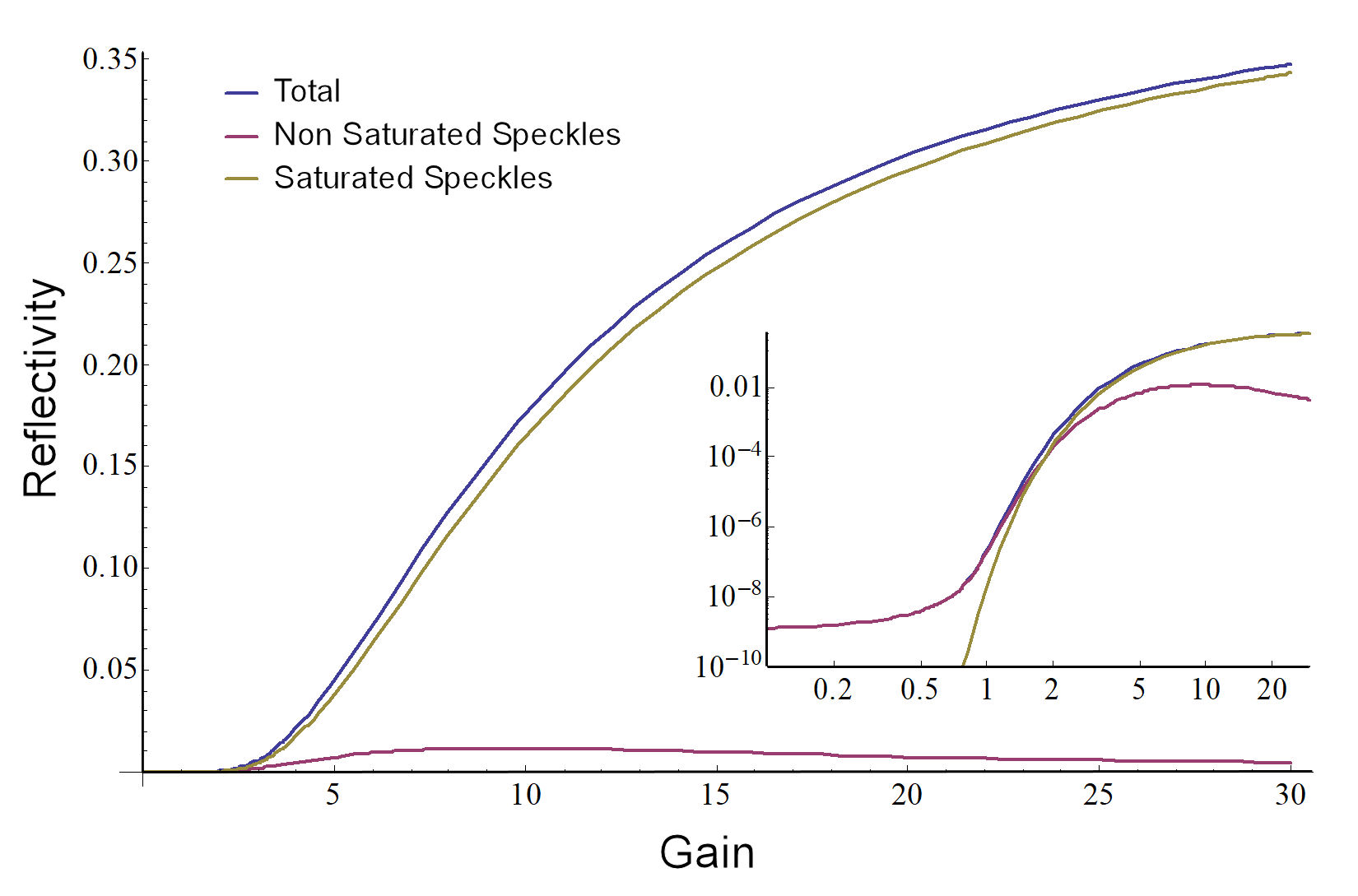}
\caption{\label{model} Values of total SRS reflectivity (blue) calculated by using Eq.(\ref{eq_refl}). Red and Yellow lines indicate the contributions given by non saturated (first term in Eq.(\ref{eq_refl})) and saturated (second term in Eq.(\ref{eq_refl})) speckles. The inset represents the graph in Logaritmic scale.} 
\end{figure}

As shown in Fig.(\ref{model}), the SRS backscattering is strongly dominated by the high intensity tail of speckle distribution, i.e. by the second term in Eq.(\ref{eq_refl}). By considering the simplified probability density $f(u)=e^{-u}$ for the speckle peak intensities (see \cite{rose93,HullerLPB2010}) and a saturation value of $R_{sat}=0.4$, the contribution from saturated speckles becomes dominant already for values $g_0 > 2$. At $g_0=2$ the speckles with $u>6$ contribute mostly to the backscattering, which is however, not yet saturated. For $g_0=5$ already the speckle population with $u\geq 4$ dominates, yielding already 10\% backscatter.

The practical expression for the convective SRS gain $g$ for the scattered light intensity, is given by\cite{Pesme}
\begin{equation}
g = 7.6 \ I_{16} \left(\frac{\lambda_0}{0.527 \mu m}\right)^2 \  \frac{L_{\nabla}}{100\mu m} \frac{(k_{epw}/2k_0)^2}{k_s/k_0} \ , \label{gain}
\end{equation}
with $\lambda_0$ denoting the laser wavelength, $L_{\nabla}$ the density gradient length, $I_{16}$ the laser beam average intensity in units of $10^{16}$ W/cm$^2$, and $k_0$, $k_s$, and $k_{epw}$ are the wave numbers of the laser light, the scattered light, and the plasma wave,
with $k_s/k_0= (1-2\sqrt{n_e/n_c})^{1/2}$, $k_{epw}/k_0 = (1-n/n_c)^{1/2}+(1-2\sqrt{n/n_c})^{1/2}$ for backscatter.

In the shots where only the interaction beam was used, relying on the nominal laser intensity and on the density scalelength $L_{\nabla} \approx 30$ $\mu$m given by hydrodynamic simulations, we obtain a Rosenbluth gain $g \approx 2.5$, depending on the density, i.e. well below the SRS threshold, usually taken as $g_{th}=2 \pi$. Applying the multispeckle model introduced above, it results in a reflectivity of the order of 0.1 $\%$, which is below the detection threshold of SRS in our experimental setup ($R_{th} \approx 0.5 \%$). This explains the lack of SRS detection in these shots. 

When heating beams are used, the density scalelength $L_{\nabla}$ increases with the delay time $\Delta \tau$ between heating and interaction beams, ranging from 150 $\mu m$ to 450 $\mu m$. By taking $L_{\nabla}=400\ \mu m$ at $n_e/n_c=$ 0.05, the gain obtained for $I_{16}=$ 1 is very high, $g\simeq 30$, in a fully saturated regime.
The gain decreases, however, to $g\simeq 21$ if we account for the local laser intensity $I\approx 7 \cdot 10^{15}$ W/cm$^2$ in the region $n = 0.05\ n_c$; this is due to the larger laser spot at large distances from the target.
Furthermore, the strong Landau damping $\gamma_L$ of the EPW significantly reduces the spatial growth rate, decreasing as $\gamma_0^2 /\nu_s \gamma_L$, where $\gamma_0$ and $\nu_s$ represent the homogeneous SRS growth rate and the group velocity of the scattered wave. According to classical convective theory, however, the reduction of the spatial growth rate is compensated by the corresponding increase of the amplification length\cite{Williams}, rising as $\gamma_L / \kappa^{\prime}\nu_e$, where $\kappa^{\prime}$ and $\nu_e$ represent the spatial derivative of the wavenumber mismatch of the 3-wave coupling and the group velocity of the EPW, respectively. Therefore, even in strong Landau damping regime, the resulting total gain coincides with the undamped standard Rosenbluth expression $g=2 \pi \gamma_0^2 / \kappa^{\prime}|\nu_e\nu_s|$, from which Eq.(\ref{gain}) is obtained, meaning that Landau damping does not affect the scattered SRS light.
In our conditions, however, the extension of the amplification length is limited by the length of the speckle $l_{\parallel}$ where SRS is driven, which implies that the gain expressed in Eq.(\ref{gain}) can be considered valid only when $(\gamma_L/\omega_{epw}) L_{\nabla} < l_{\parallel}/2 $. If the condition is not verified, the Rosenbluth-type amplification has to be replaced by spatial amplification in a limited homogeneous plasma, where the growth rate depends on the plasma wave damping $\gamma_L$.
By taking $L_{\nabla}= 400 \ \mu m$, the latter condition is verified for a damping rate $\gamma_L/\omega_{epw} >$ 0.8\% which is met for plasma densities $n_e/n_c <$ 0.07 at $T_e\sim$ 1 keV. 
Practically, the general expression for the gain that should be applied for spatial amplification in Eq. (\ref{eq_refl}) can be written as
\begin{equation}
g =  0.27 \ I_{16} \ \lambda_{0}^2 \ \frac{(k_{epw}/2k_0)^2}{k_s/k_0} \ \min\{ L_{\nabla}\ , \frac{\omega_{epw}}{4 \gamma_L}\frac{k_{epw}}{k_0}l_{\parallel} \} \label{gain_gen}
\end{equation}
where $\lambda_0$, $L_{\nabla}$, and $l_{\parallel}$ are expressed in units of $\mu$m.

Via the latter expression we can evaluate the gain in individual laser speckles.
For $L_{\nabla}= 400\ \mu m$, $n_e/n_c=$ 0.05 and $I_{16}=$ 0.7, assuming that linear Landau damping applies, we obtain a reduced gain in the range $g=7-10$ for plasma temperatures T = 1 - 1.2 keV. 

\begin{figure}
\includegraphics[width=0.48\textwidth]{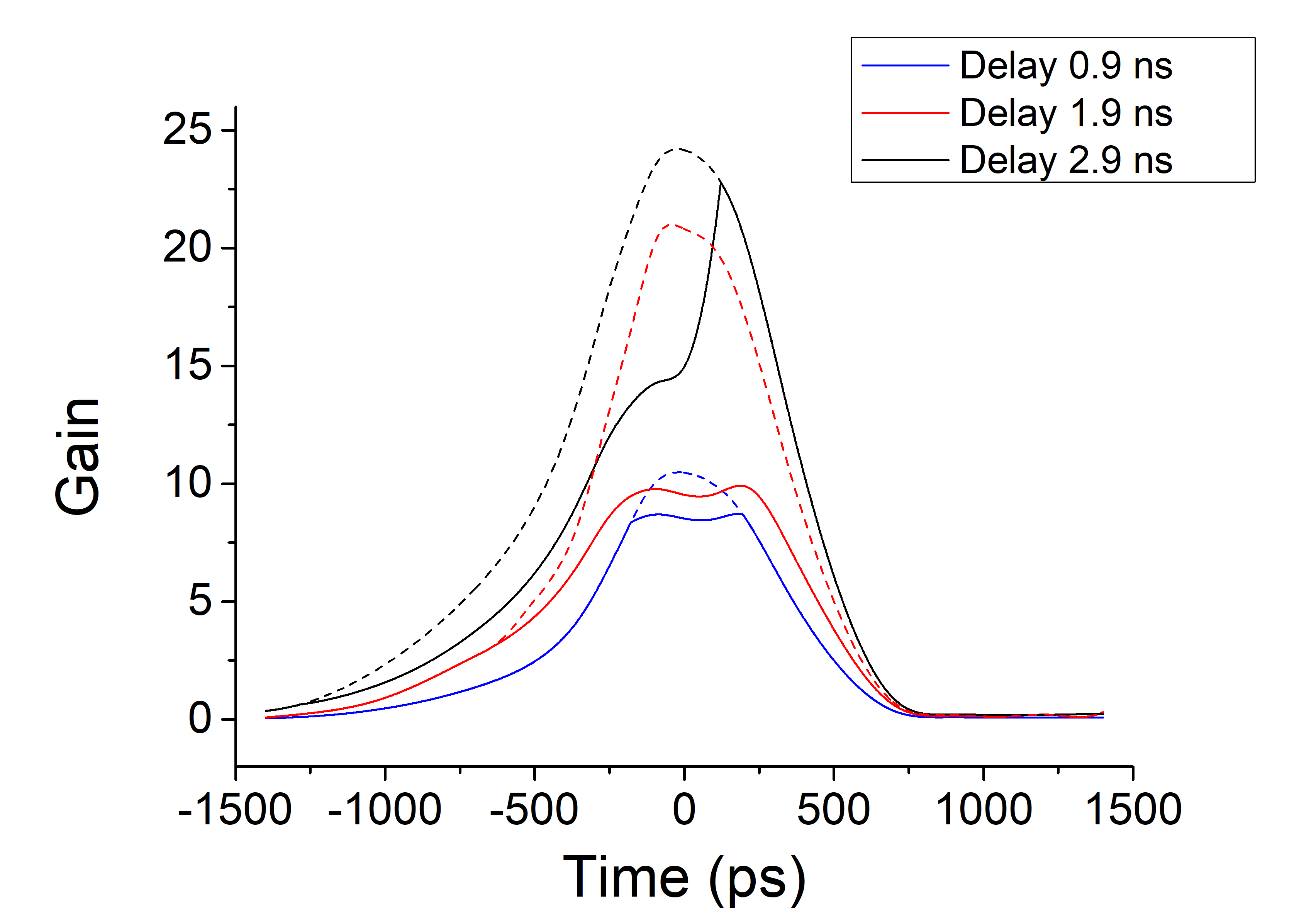}
\caption{\label{model_profile}Time Evolution of SRS gain calculated by using Eq.(\ref{gain_gen}) for the laser beam configurations with time delay $\Delta\tau=$ 0.9, 1.9 and 2.9 ns between heating and interaction beams. Dashed and solid lines indicate the gains calculated by using the Rosenbluth theory and the modified expression Eq.(\ref{gain_gen}) accounting for Landau damping into a speckle, respectively.}
\end{figure}

%
It is instructive to calculate the SRS gain along the time profile of the laser pulse by using the model described above, applying Eq.(\ref{gain_gen}).
In Fig. (\ref{model_profile}), the time evolution of the SRS gain at $n_e / n_c$ = 0.05 is reported for the cases where the time delay $\Delta \tau=$ 0.9, 1.9 and 2.9 ns. Values of density scalelength and temperature are here taken from hydrodynamic simulations. When $\Delta\tau$ increases from 0.9 ns to 2.9 ns, $L_{\nabla}$ rises from 200 $\mu$m to 450 $\mu$m; however, since the plasma coronal region becomes longer with time, the region of density $n_e / n_c$ = 0.05 moves farther from the ablation region, leading to a slight decrease of temperature with $\Delta\tau$.
For each delay time, the pure Rosenbluth gain, where the amplification length is not limited by the extension of the speckle, is reported as a dashed curve. The real gain (solid lines), however, is lower than this curve at some times, where Landau damping is strong and the second term in Eq.(\ref{gain_gen}) becomes dominant. For the lowest value $\Delta \tau=$ 0.9 ns, Rosenbluth gain prevails at all times - except at times close to the laser peak - because of the steep profile of the plasma. Differently, for the highest value $\Delta \tau=$ 2.9 ns, the Rosenbluth gain prevails only in the trailing part of the main pulse, where the heating beams are switched off and the plasma temperature rapidly falls, reducing the Landau damping. Finally, in the middle case $\Delta \tau=$ 1.9 ns, the gain is strongly dominated by Landau damping at all times.

By applying the model from Eq.(\ref{eq_refl}) with $R_{sat}=$ 0.3 / 0.4 / 0.5, respectively, and integrating along the pulse profile, the gain values plotted in Fig.(\ref{model_profile}) result in an overall reflectivity of
8\% / 9\% / 11\% for $\Delta\tau=$ 0.9 ns, 10\% / 11\% / 16\% for $\Delta\tau=$ 1.9 ns, and 17\% / 22\% / 27\% for $\Delta \tau=$ 2.9 ns.
Although the gain values from Eq.(\ref{gain_gen}) indicate lower bound values, since possible effects due to non linear Landau damping (kinetic effects) are not taken into account, the reflectivity values which are obtained are in a good agreement with the experimental data shown in Fig.\ref{SRS}b, also considering the uncertainties on the plasma temperature and on the value of $R_{sat}$. Further, Fig.(\ref{model_profile}) shows that the model qualitatively reproduces the increase of SRS duration with $\Delta\tau$ observed in the experiment (see Fig.\ref{SRS}d). 
\subsubsection{Validation of the model and role of kinetic effects}
In order to validate the model described above and to investigate the relevance of kinetic effects on the SRS growth, we have performed 2D and 3D simulations with the wave-coupling code SIERA\cite{Tran} (CEA and CPHT), computing the plasma response of an optically smoothed laser beam. Simulations consider a linear density ramp in the range  0.02 $\leq n_e/n_c \leq$ 0.10, over a propagation length of several hundred wavelengths and several speckle lengths, concretely a simulation volume of 600 x 100 x 100 $\mu m^3$, along and across the laser propagation axis, respectively.
They describe the SRS growth in an isothermal plasma from an ensemble of about 1000 speckles for the duration of 10-20 ps, corresponding to the peak of the laser pulse.
The modular concept of the SIERA code allows us to take into account, or not, kinetic effects due to trapped particles excited by high amplitude EPWs, which result in a departure from the Landau damping and in a detuning of the SRS resonance condition\cite{MoralesONeil,Dewar,DewarLindl,RoseRussellPoP2001,Strozzi2007}. In previous studies, in particular, it was shown that in inhomogeneous plasmas the spatial amplification in speckles can be destabilized by the effect of auto resonance due to the generation of trapped electrons, eventually leading to higher SRS backscatter\cite{Tran,chapmanPRL}. 

While the model described in the previous section yields already a satisfactory agreement with the experimental results in terms of SRS backscattered energy, numerical simulations reveal further features:
(i), the spectral width of the backscattered light is found in the window roughly in between 650 nm and 750 nm, in agreement with the experiment; (ii) SRS exhibits an overall bursty behaviour in the spectral emission.

\begin{figure}[tb]
\includegraphics[width=0.233\textwidth]{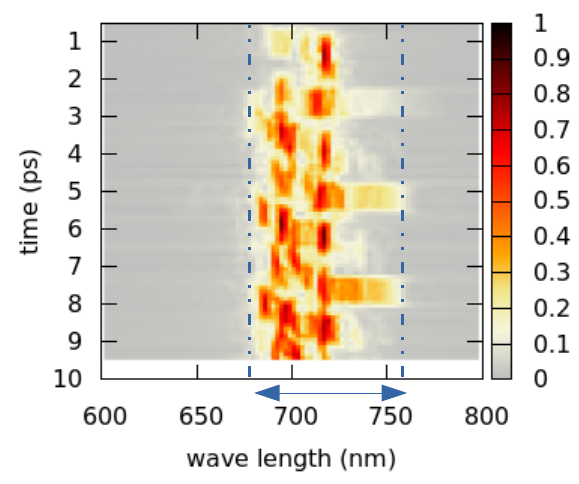}
\includegraphics[width=0.239\textwidth]{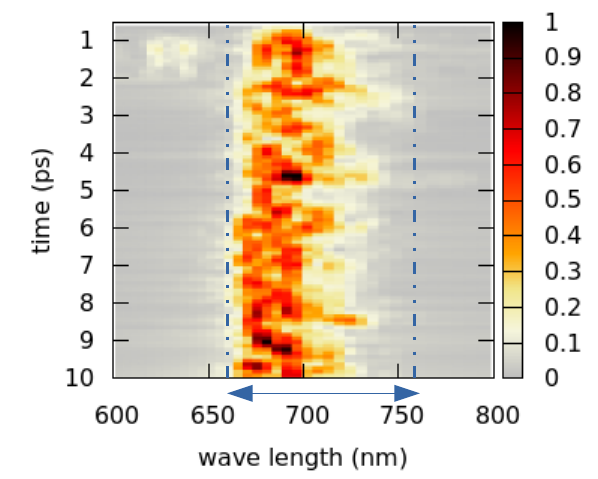}
\caption{Spectra of the backscattered light from simulations with SIERA for a RPP beam at $I_0=10^{16}$ W/cm$^2$ and for a density ramp 0.02 $< n/n_c <$ 0.10, left/right subplots without/with taking into account kinetic effects.\label{siera-spectrum}}
\end{figure}

SIERA simulations (in agreement with the PIC simulations described in the next section) show that kinetic effects in intense speckles affects the SRS light spectrum. The excitation of trapped electrons leads in fact to a broader SRS spectrum and, in addition, produces a shift towards lower EPW frequencies, thus towards shorter wavelengths in the scattered light, in the range 650-700 nm, as shown in Fig. \ref{siera-spectrum}.
The bursty behaviour in the simulations occurs on a very short, ps-time scale, which is not fully resolved in the experiments. The bursts reflect the amplification of the scattered light over the dominating density range of SRS amplification, and disruptions due to transient pump depletion in intense speckles.
%

The time-average of the bursty behaviour in the multiple-speckle simulations yield SRS reflectivity values of $\langle R\rangle \sim$ 0.4-0.45 for both the cases with and without kinetic effects for $I_0=$ 10$^{16}$ W/cm$^2$; this value is in good agreement with the model presented above by taking $R_{sat}\approx 0.5$. Simulations show the negligibility of kinetic effects in determining the SRS amplification at this high laser intensity. Vice versa, they reveal that kinetic effects produce a strong boost of SRS reflectivity at lower laser intensities, which are closer to the SRS threshold. This explains the quite large reflectivity $\langle R\rangle\sim$0.3 obtained with SIERA simulations at the much lower intensity $I_0=3.5\cdot$10$^{15}$ W/cm$^2$.

These results also suggest that the model described in the previous section is adequate at the high laser intensities used in this experiment, where the role of kinetic effects on reflectivity is marginal. For lower average beam intensity the situation can be different.
In the model of Eq.(\ref{eq_refl}) we have assumed the SRS gain from Eq.(\ref{gain_gen}) by evaluating the linear Landau damping value. Our simulations, and those in other work \cite{Yin_Pop2012, YinPoP2013} suggest that hot electrons emerging from high-amplitude EPWs in intense speckles massively modify the distribution function of neighbouring lower intensity speckles. In this case the local and most likely lower value of Landau damping, as well as the EPW frequency shift, determined from the local electron distribution, need to be considered. The latter may favour a more vigorous onset of SRS even in lower intensity speckles, such that our model would underestimate the SRS reflectivity for lower average beam intensity.

\begin{figure}
\includegraphics[width=0.45\textwidth]{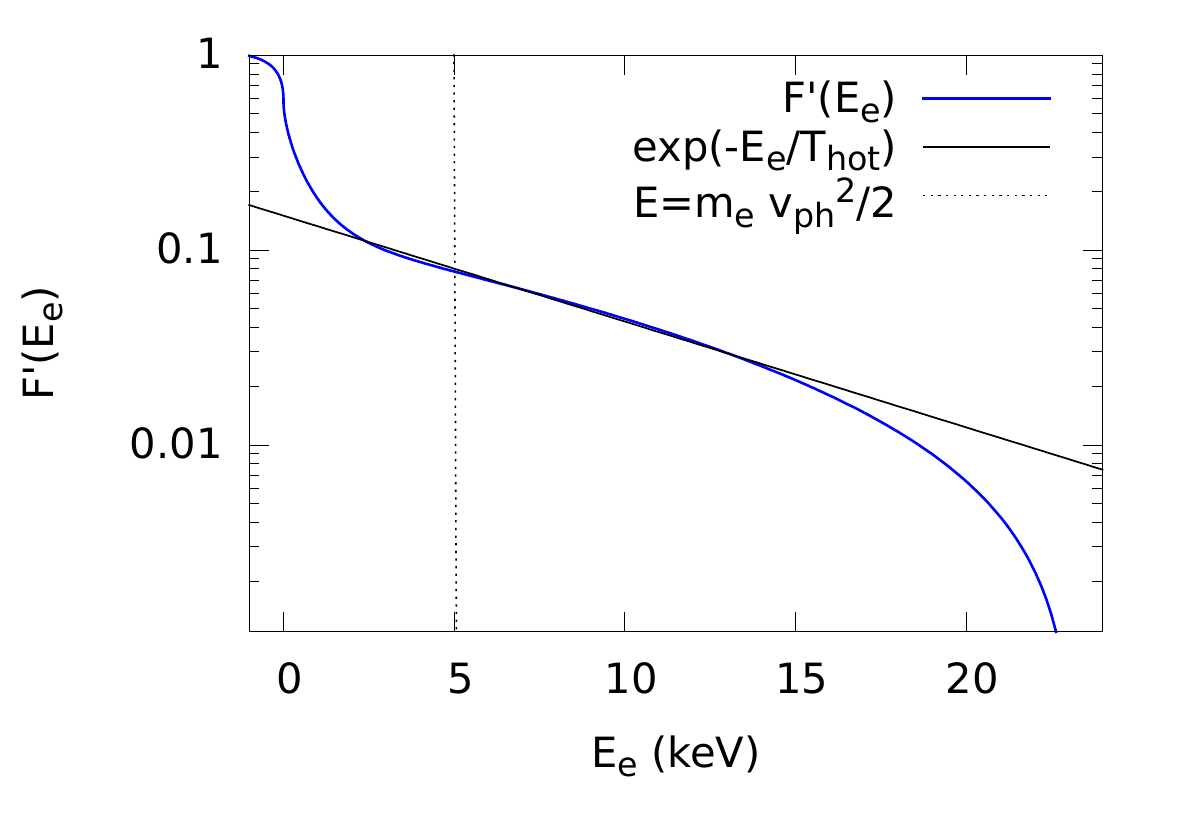}
\vspace{-.5cm}
\caption{Complementary distribution function $F'(E_e)$ as a function (blue line) of the electron energy $E_e$ from a typical simulation with SIERA for a RPP beam at $I_0=10^{16}$ W/cm$^2$ and for a density ramp 0.02 $< n/n_c <$ 0.10. The tail of the distribution evolves versus a HE distribution (black line)  $\sim \exp(-E_e/T_{hot})$ with $T_{hot}=$ 8 keV here, i.e. in the range of the experimentally observed values. The position of the energy corresponding to the EPW phase velocity is indicated (dotted line at $\sim$ 5 keV).\label{energy_spectrum}}
\end{figure}

SIERA simulations with kinetic effects also depict the evolution of the electron energy distribution function via the EPW model used in the code\cite{Tran}. 
The model for non linear EPWs is here certainly incomplete with respect to more thorough approaches concerning particle trapping\cite{RoseRussellPoP2001}.
However, in the frame of validity of the model, it is possible to deduce the complementary electron distribution function $F'(E)\equiv \int_E^{\infty} f(E') dE'$ which is shown in Fig. \ref{energy_spectrum}; it exhibits the form of a hot tail beyond the energy value $E_e(v_{ph})\simeq$ 5 keV corresponding to the EPW phase velocity $v_{ph}\sim$ 0.14 $c$, to which a temperature $T_{hot}$ in the range 8-11 keV can be associated. 
This value is consistent with the experimental measurements. 

\subsubsection{Self focusing in intense laser speckles}
In order to analyze more in detail the physics in single intense speckles we have also carried out simulations with the 2D3V Particle-In-Cell (PIC) code EMI2d of CPHT\cite{EMI2d-2004,EMI2D-2010,riconda2011}. 

As mentioned earlier, the optically smoothed laser beams used in the experiments contain speckles with up to 8-10 times the average beam intensity $\langle I \rangle$, and, for the current plasma, speckles with $I_{sp} > 3.5 \langle I\rangle$ will be subject to self-focusing. In these speckles, a significant density depletion is expected to give rise to a reduction or even a suppression of stimulated scattering in the center of the hot spot, as shown in Ref. \cite{Masson-Laborde2014}. 

In the simulations carried out with the EMI2d code we focussed on the micro-physics around an intense speckle with $v_{\rm osc}/c =$ 0.1 corresponding to a peak intensity of $5\cdot 10^{16}$ W/cm$^2$ for the conditions considered.
Simulations have been carried out in an inhomogeneous density profile with $L_{\nabla}=$ 400 $\rm\mu m$ around $n_e=$ 0.04 $n_c$.
They evidently show kinetic effects visible both in the frequency spectra, as modeled in SIERA, and in the HE spectrum; for the mentioned simulation with a high intensity speckle a HE temperature of 8 keV has been determined, with which the result from the SIERA simulation is consistent.

\begin{figure}
\includegraphics[width=0.48\textwidth]{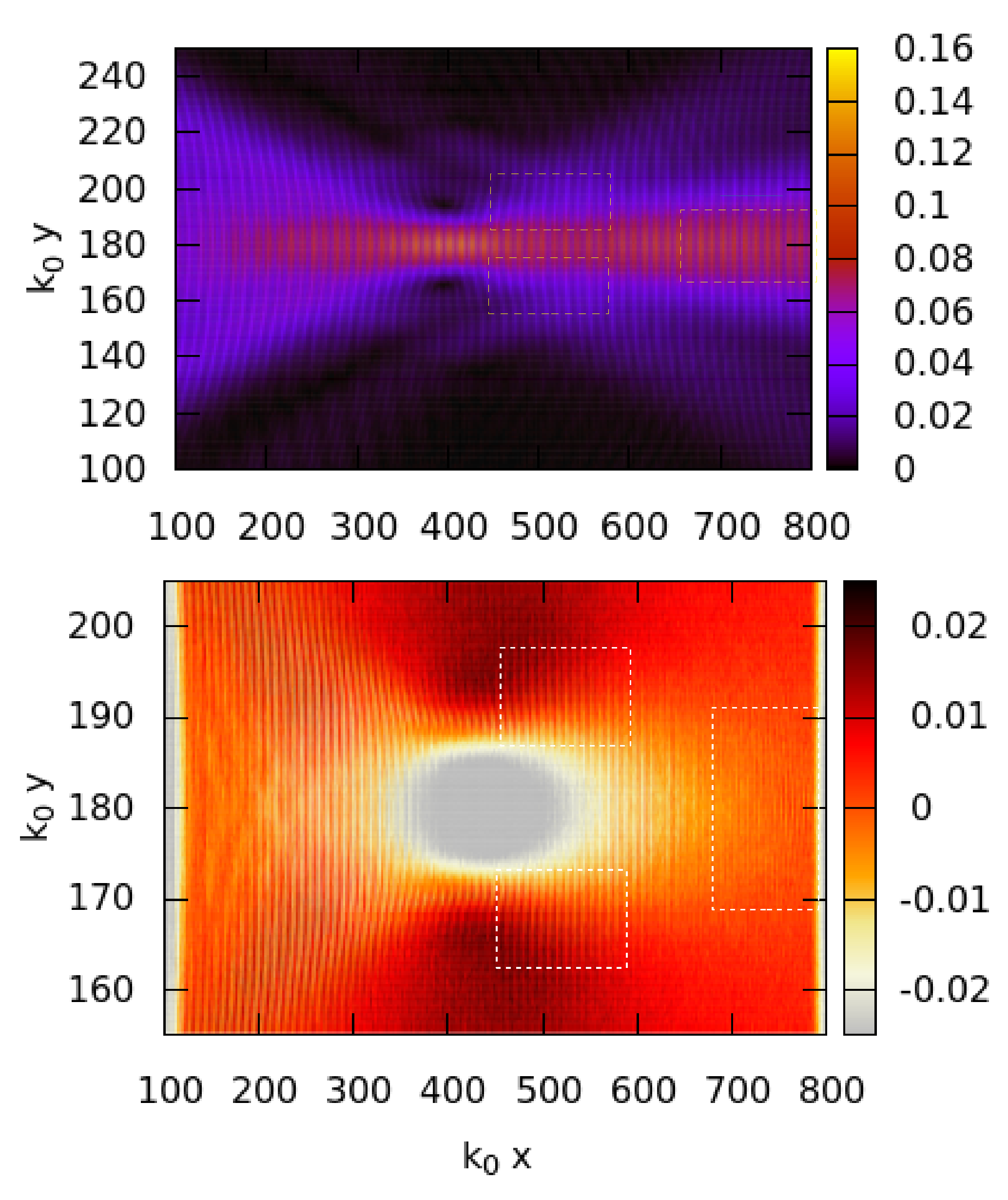}
\vspace{-.4cm}
\caption{\label{pic-snapshots} Normalized transverse field strength $e|E_y|/(m_e \omega_0 c)$ (upper subplot) and electron density perturbations (lower subplot) $(n_e(x,y)-n_0(x))/n_0(x)$ as a function of space for a single laser speckle with peak field strength $e|E_y|_{in}/(m_e \omega_0 c )=$0.1. The self-focusing speckle shows a elongated structure on axis ($k_0 y=$180) with a field strength higher than the initial value $|E_y|_{in}$.
The laser light enters from the left, $x=$0.
The dashed lines in both subplots indicate the zones where SRS is amplified:
the density perturbations (note that the range shown in $y$ is reduced with respect to the upper subplot) exhibit plasma wave oscillations in the periphery of the field hot spot, but not in the depleted zone inside. SRS-driven EPW are in particular present at the rear on axis and in side wings. SBS-driven ion acoustic wave are found in the front part (PIC simulations show generally too high values).}
\end{figure}

EMI2d simulations were carried out with and without mobile ions, in order to discriminate the impact of ponderomotive self-focusing.
As expected, the self-focusing provokes a local density depletion in the speckle, such that both back-scattering instabilities (SRS and SBS) develop preferentially in the periphery of the speckle, but not in its center, as shown in Figs.\ref{pic-snapshots}. 

Here, however, a clear distinction is visible between the actions of SBS and SRS, which is due to their different time-scale. While SBS grows on times of the same order as the self-focusing, SRS has a much faster growth\cite{CohenPoP2001,riconda2011}, allowing the SRS to adapt very rapidly to ponderomotively induced profile modifications. In this way, SRS (re-)establishes rapidly in the arising filamentary structure, preferentially in the rear of the self-focused speckle in which the plasma density is not (yet) depleted. SRS therefore continues to contribute vigorously to the backscatter process and leads eventually to higher backscatter rates than without self-focusing, as already reported in previous experiments\cite{Montgomery}. The limited capacity of SRS mitigation by optically smoothed laser beams was demonstrated by Ferndandez et al. \cite{FernandezPoP2000}.

Let us mention that SBS is also seen in our PIC simulations with mobile ions, but starts from an unnaturally high noise level because of the limited number of particles; in contrast to SRS, however, SBS is mitigated by the ponderomotively depleted density profile\cite{Masson-Laborde2014}.

\subsection{SRS at higher densities}
In order to investigate the effect of pump depletion and filamentation driven at low densities on the laser light propagation in more dense regions in the plasma, we carried out further simulations where a SIERA module was implemented into the hydrocode HERA\cite{LoiseauPRL,myatt2001}. They confirmed a strong self-focussing in more intense speckles and showed a vigorous growth of SRS primarily in the first 2-3 speckles layer irradiated by the beam. Beyond this zone, as already mentioned earlier, the formation of temporarily incoherent filaments, due to speckles self-focusing, arises and provokes the so-called "dancing filaments" \cite{SchmittAfeyan} and the onset of plasma-induced smoothing\cite{maximov2001,labaune2000,LoiseauPRL} further inside the plasma.
This has consequences on the LPI driven in denser regions and explains the weak SRS observed at higher plasma density, which is discussed in this section.

The origin and the time-evolution of the weak SRS observed at densities around $0.10$ n$_c$ are completely different from those described in the previous sections. Here, SRS appears at late times of interaction at densities around $n_e=0.13$ $n_c$, corresponding to $k_{epw}\lambda_D\approx0.23$, and fades after a few hundreds ps at densities around $n_e=0.08$ $n_c$, corresponding to $k_{epw}\lambda_D\approx0.30$. Scattered light is here partly reabsorbed by collisions on its way out of the plasma, which we estimated to amount to $\approx 20\%$ re-absorption for light originating at $n_e=0.1$ $n_c$ and to $\approx 30\%$ for light originating at $n_e=0.15$ $n_c$. These values, however, are not so large to support the hypothesis that SRS could be driven at densities higher than $n_e=0.13$ $n_c$ and not be observed because of light re-absorption.

Looking at the longitudinal profiles of density given by the hydro simulations (Fig. \ref{DUED}e), it is evident that the densities of interest correspond to the region of the cavity drilled by the interaction beam. The density at the bottom of the cavity decreases from $n_e\approx0.2$ $n_c$ to $n_e\approx0.08$ $n_c$ for the times corresponding to SRS emission from higher densities. This explains the temporal shift of SRS spectrum observed in the experiment.
The strong density perturbations observed at even larger distances from the target, and the temporal evolution in these regions exclude that SRS could be driven there.

The density in a cavity can be described by a parabolic profile  $\omega^2_p(x)=\omega^2_{p0}(1+x^2/L_{\nabla}^2)$, where $L_{\nabla}$ is the density scalelength and determines the resonance length $l_{res}$ for SRS.
Absolute SRS can temporally grow at the minimum density if $l_{res}$ exceeds the homogeneous growth length and if the damping of the daughter waves can be overcome. The bottom of the cavity represents a point of phase inflection where the wavenumber mismatch $\kappa=k_0\!-\!k_s\!-\!k_{epw}$ can be expressed by $\kappa(x)=\kappa^{\prime\prime}(0)x^{2}/2$ where $\kappa(0)=0$ and $\kappa^{\prime}(0)=0$. Here, the parabolic density profile allows the electron plasma waves to be trapped into the cavity and to be described as solutions of a quantum harmonic oscillator, as suggested by Barr et al.\cite{Barr}. Since the SRS homogeneous growth rate $\gamma_0=5.7\cdot10^{-3}\omega_0$ is here larger than the separation of the frequency modes $\Delta\omega\approx v_e/L_{\nabla}\approx9.5\cdot10^{-5}\omega_0$, however, the quantized mode structure is washed up and the classical WKB approach of absolute SRS growth at a density "extremum" can be applied.
Here, the SRS threshold for the most unstable mode, corresponding to the onset of the instability at the bottom of the cavity, can be expressed by $\Gamma^2>0.15$ where $\Gamma^2=\gamma_0^2/\nu_e \nu_s (\kappa^{\prime\prime})^{2/3}$ and $\kappa^{\prime\prime}=\omega^2_p/3k_{epw} v^2_e L_{\nabla}^2$. In cases where the instability growth dominates the damping of the daughter waves, i.e. $\gamma_0>(\sqrt{\nu_e\nu_s}/2)(\gamma_L/\nu_e + \gamma_{coll}/ \nu_s)$, SRS grows in the absolute regime, while in case of non negligible damping SRS can still spatially grow in the convective regime \cite{Williams2}.

In our case, a scalelength value $L_{\nabla}\approx45\,\mu m$ can be estimated by fitting the density profile obtained from hydrosimulations. According to the relation $l_{res}=(9v^2_e L_{\nabla}^2 k_{epw}/\omega^2_p)^{1/3}$, reported in Ref.\cite{Afsharrad}, we can estimate a value of the resonance length $l_{res}=4\,\mu m$, much shorter that the length of a laser speckle.
By considering the nominal laser intensities at relevant times, the $\Gamma^2$ value decreases from 0.51 to 0.31 during the times of SRS observation, due to the fall of laser intensity at late times of interaction. The SRS growth rate exceeds the Landau damping value of plasma waves - dominating on collisional damping of scattered light - when SRS is driven at $n_e=0.13\,n_c$ ($k_{epw}\lambda_D=0.23$), while damping becomes stronger when the cavity becomes too deep ($k_{epw}\lambda_D=0.30$) halting the instability.
Local values of laser intensity can be however significantly different from nominal ones: on one side, they are strongly reduced by pump depletion of laser light suggested by the measured SRS reflectivity, by the collisional absorption of laser light and by the 'dancing filaments' effect, as discussed above, which could explain the weak SRS signal observed; on the other side, again, intensity statistics in speckles can play a major role in the SRS onset.
\subsection{SUMMARY}
In the present experiment we characterized laser interaction with a long plasma corona with density scalelengths up to 450 $\mu m$ and intensities at $\lambda_0 = 0.527\ \mu m$ up $2\cdot 10^{16}$ W/cm$^2$, i.e. close to values typical of Shock Ignition scheme of ICF.
In addition to a considerable reflection of light at wavelengths $\approx \lambda_0$ ($15-35\ \%$), including a non quantified amount of SBS backscatter, the experiment showed a large SRS backscatter ($4-20\ \%$), increasing with the scalelength of the plasma, at low plasma densities $n_e \approx \ 0.05\ n_c$, in a region prone to strong Landau damping ($k_{epw}\lambda_D =0.3-0.5$). Occasionally, a very weak SRS signal is measured at higher densities, while no signatures of LPI a quarter critical density is observed, confirming that non-collisional coupling is driven only in the low dense plasma corona.
The values of SRS reflectivity obtained with different plasma gradients are well reproduced by a simplified model, accounting for the contribution given by independent speckles, with local intensities following an exponential distribution, and for the reduced gain due to classical Landau damping. The model clearly shows that the reflectivity is dominated by SRS driven in speckles in a saturated regime.
In order to investigate the micro-physics resulting in SRS saturation into the speckles, and to evaluate the role of kinetic effects and speckle self focussing, simulations were carried out by using the wave-coupling code SIERA and the Particle-In-Cell code EMI2d. It was shown that SRS results in a general bursty behaviour by transient pump depletion, which is mainly responsible for SRS saturation. In addition, kinetic effects result in a broadening and in a blue shift backscattered light spectrum. No clear enhancement of SRS due to kinetic effect was however shown in the simulations, which is due to the high laser intensity. PIC simulations showed the occurrence of self focussing in the more intense speckles, leading to an enhancement of SRS, which is able to rapidly adapt to the modified density profile, due to its growth rate. Filamentation in high-intensity speckles also suggests the occurrence of plasma-induced smoothing after a few speckle layers, reducing the coherence of light reaching higher density regions and explaining, along with pump depletion, the lack of SRS/TPD in these regions.
Finally, the experiment reveals the generation of low-energy HE ($T_{hot}\sim 7-12$ keV). These electrons are probably produced by a multiplicity of mechanisms, including the acceleration in EPWs produced by SRS at low plasma densities, as suggested by SIERA and EMI2d PIC simulations, and can be explained by the low phase velocity of plasma waves in this region.

\section{CONCLUSIONS}

Results described in this paper confirm the importance of speckle pattern and intensity distribution in determining local conditions of interaction and parametric instabilities growth. The long scale plasma and the high laser intensity, driving self focussing of speckles, produce strong SRS amplification at low plasma densities. Even though the higher predicted plasma temperatures of full-scale SI are expected to enhance kinetic effects and to shift SRS outburst at slightly higher densities, present results suggest that pump depletion and plasma smoothing prevent or at least strongly decrease the impact of non collisional LPI at quarter critical density, consequently limiting the generation of electrons with energies able to preheat the fuel capsule is inhibited. 
Overall, our results suggest that a good control of both filamentation and SRS remains a key factor for the success of the Shock Ignition scheme. Plasma smoothing induced by filamentation could be a route to produce a beneficial incoeherence on the beam, needed to prevent the loss of laser energy in low density plasma regions via non collisional processes. However, a reduction of SRS driven in regions where plasma smoothing is still inactive, which could be obtained by using large bandwidth lasers, is required to avoid the large SRS reflectivity, probably reaching 40-50$\%$ in the full angular range.

\appendix\label{Appendix}
\section{APPENDIX}
\subsection{Model for multiple speckle backscatter}
We consider an optically smoothed laser beam formed by numerous speckles of equal size generated by a Random Phase Plate for which the peak intensity $I_{sp}$ of the speckles follows a probability distribution $f(I_{sp})$.
The so-called independent speckle model \cite{rose93} is based on the fact that each laser speckle contributes incoherently to the total power (energy) of the backscattered light from stimulated scattering processes, like SBS or SRS.
The backscattered light flux $S_{RPP}$ from a laser beam generated by a RPP is then essentially the sum of the contributions $S_{j}$ of the $n_{sp}$ individual speckles,
\begin{equation}
S_{RPP} = \sum_{j=1}^{n_{sp}} S_j \equiv R_{sp, j}(I_{sp,j}) I_{sp,j}\ , \label{eq2} 
\end{equation}
in which $S_j$ is expressed as the backscattered fraction $0\leq R \leq 1$ for the $j$-th speckle with its peak intensity $I_{sp,j}$. 
While the speckle field of a finite number of speckles has statistical fluctuations from the average probability density $f(I_{sp})$ for each RPP realisation\cite{HullerLPB2010}, once $f(I_{sp})$ is known, one can determine the average backscattering via the integral
\begin{equation}
S_{RPP} = \langle R\rangle \langle I\rangle = \int_{0}^{I_{max}} R_{sp}(I)\ I\ f(I) dI \ , \label{eq3}
\end{equation}
with $\langle I\rangle$ denoting the average speckle intensity, $\equiv  \int_{0}^{I_{max}} I\ f(I) dI$ and with $\int_{0}^{I_{max}} f(I)  \equiv 1$. Since $\langle I\rangle = I_0 $, where $I_0$ is the envelope laser intensity of the beam, $\langle R\rangle$ corresponds to the total reflectivity which is experimentally measured. The intensity of the most intense speckle, $I_{max}$ is a function of the number $n_{sp}$ of speckles in the interaction volume and can be expressed by 
$I_{max}/\langle I\rangle \simeq \log n_{sp} +\gamma_E$ (with $\gamma_E\simeq$.57724) with an important spread\cite{HullerLPB2010}.

The response function of each speckle 
splits in two regimes (i) and (ii), 
namely (i) for `lower-intensity' speckles with strong dependence on $I_{sp}$ up to the saturated regime for which (ii) the response on the speckle intensity is only very weak, 
such that  Eq.(\ref{eq3}) simplifies to
\begin{equation}
\langle R\rangle \langle I\rangle \!\simeq\!\! \int_{0}^{I_{sat}}\!\! \!R_{sp}(I)\ I\ f(I) dI +\! R_{sat}\! \int_{I_{sat}}^{I_{max}}\!\! I\ f(I) dI  .\label{eq4}
\end{equation}
The physics of the saturated regime is simplified in this model by the ensemble- and time averaged saturation rate of the backscattering, $R_{sat}$. See herefore the discussion in the main text.

Following the classical theory formulated by Rosenbluth\cite{Rosenbluth}, the growth of stimulated scattering in an inhomogeneous plasma, here SRS, can be expressed by $I_{SRS}=I_{noise}\exp(g)$ where the amplification gain $g=2 \pi \gamma_0^2 / \kappa^{\prime}|\nu_e\nu_s|$ expresses the dependence on the laser intensity, via the standard SRS growth rate $\gamma_0 \propto \sqrt{I}$, and on the density gradient, via the spatial derivative $\kappa^{\prime} \propto L_{\nabla}^{-1}$ of the wavenumber mismatch of the 3-wave coupling, $\kappa=k_0\!-\!k_s\!-\!k_{epw}$, with $L_{\nabla}$ denoting the density scale length in the plasma.

The SRS reflectivity in a speckle of intensity $I = u\ \langle I\rangle$, i.e.,
$I_{sp}^{SRS}(u)  =  R_{sp} u\ \langle I\rangle$ 
can then be evaluated 
via the amplification gain, $g_0$, of a speckle at average intensity
\begin{equation}
R_{sp}(u)  = \varepsilon\ e^{g_0 u},  \label{eq6}
\end{equation}
%
where $\varepsilon = I_{noise} / u \langle I\rangle$ stands for the ratio of the noise level to the speckle intensity, whose typical value for warm plasmas in laser plasma interaction is roughly $10^{-9}$. The speckle reflectivity $R_{sp}$ saturates, due to depletion of the incoming flux or due to non linear effects in the plasma wave, at a level $R_{sat}$ for the speckle intensity $I_{sat} = u_{sat}\ \langle I\rangle$, which implies that the intensity of the speckle intensity for which saturation occurs, $u_{sat}= g_0 \log (R_{sat}/\varepsilon) \simeq (20 -\log R_{sat})/g_0$ depends on the gain $g_0$ and only weakly (logarithmically) on the saturation value $R_{sat}$. 
Evidently $R_{sp} = R_{sat}$ for $u \geq u_{sat}(g_0,\log R_{sat})$.
%
With Eq.(\ref{eq6}) in Eq.(\ref{eq4}) one obtains
%
\begin{equation}
\langle R \rangle= \varepsilon \int_{0}^{u_{sat}}\!\! u\ e^{g_0 u} f(u)\ du + R_{sat} \int_{u_{sat}}^{u_{max}}\!\! u\ f(u)\ du  ,\label{eq7}
\end{equation}
in which $u_{max}$, corresponding
to the most intense speckle, is much greater than unity for $n_{sp}>$1000; in practice this suggests to use $u_{max}\!\rightarrow\!\infty$ in Eq. (\ref{eq7}).
%
%
\section*{acknowledgements}
We would like to acknowledge financial support from the LASERLAB-EUROPE Access to Research Infrastructure activity within the EC’s seventh Framework Program (Application No. 18110033).
This work has also been carried out within the framework of the EUROfusion Enabling research projects  AWP19-20-ENR-IFE19.CEA-01 and AWP21-ENR-01-CEA-02, and has received funding from the Euratom research and training programme 2019-2020 and 2021-2025 under grant no. 633053. The views and opinions expressed herein do not necessarily reflect those of the European Commission.
We also acknowledges financial support from the CNR funded Italian research Network ELI-Italy (D.M. No.631 08.08.2016)
Simulations by CPHT were granted access to the French HPC resources of IDRIS under the allocation 2020-0500573 made by GENCI, France.
\bibliography{bibliography}
\end{document}